\vskip 12mm

\baselineskip=16pt

\font\twelvebf=cmbx12

\vskip 15mm
\def\[{[\![}
\def\]{]\!]}
\def\Z{{\bf Z}}
\def\C{{\bf C}}
\noindent

\def\N{{\bf N}}

\def\e{\eqno}
\def\l{\ldots}
\def\n{\noindent}
\def\1gl{gl_0(1|\infty)}
\def\2gl{gl(\infty |1|\infty)}
\def\gl{gl(1|\infty)}
\def\q{\quad}
\def\V{V([M])}
\def\G{\Gamma}
\def\a{\alpha}

\def\t{\theta}
\def\v{\varphi }
\def\p{\tau }
\def\ep{\epsilon }
\def\rl{\rangle}

\vskip 2cm

\leftskip 50pt
\n
{\twelvebf Highest weight irreducible representations of the
Lie superalgebra $\gl$ }

\vskip 1cm

\noindent
T.D. Palev$^{a)}$ and N.I. Stoilova\footnote{ $^{a)}$}{Permanent address:
Institute for Nuclear
Research and
Nuclear Energy, 1784 Sofia, Bulgaria; Electronic mails:
tpalev@inrne.bas.bg, stoilova@inrne.bas.bg}

\noindent
Abdus Salam International Centre for Theoretical Physics,
34100 Trieste, Italy

\vskip 3cm
\n
Two classes of irreducible highest weight modules of the general
linear Lie superalgebra $\gl$ are constructed. Within each module a
basis is introduced and the transformation relations of the basis
under the action of the algebra generators are written down.

\vfill\eject

\bigskip
\leftskip 0pt

\noindent
{\bf I. INTRODUCTION}

\bigskip

%In the present paper
We construct two classes of irreducible representations of the
infinite-dimensional general linear Lie superalgebra $\gl$. Both
of them are classes of highest weight representations,
corresponding to two different orderings of the basis in the
Cartan subalgebra. Related to this it is convenient to define
$\gl$ in two different, but certainly equivalent ways.  We denote
them as $\1gl$ and $\2gl$ (see the end of the Introduction for
the notation that follow).

\smallskip
{\it Definition 1. The Lie superalgebra $\1gl$ is a complex
linear space with a basis $\{e_{ij}\}_{i,j\in \N}$. The
$\Z_2$-grading on $\1gl$  is defined from the requirement that
$e_{1j}$, $e_{j1}$, $j=2,3,\l $ are  odd generators, whereas
all other generators are even.  The multiplication {\rm (}$\equiv
$the supercommutator{\rm )} $\[\;,\;\]$ on $\gl$ is a linear
extension of the relations: }

$$
\[e_{ij},e_{kl}\]
=\delta _{jk}e_{il}-
(-1)^{deg(e_{ij})deg(e_{kl})}
\delta_{il}e_{kj}, \;\; i,j,k,l\in \N. \eqno (1)
$$

\smallskip
As a basis in the Cartan subalgebra ${\cal H}_0$ we choose
$\{e_{ii}\}_{i\in \N}$ with a natural order between the
generators: $e_{ii}<e_{jj}$, if $i<j$. Then ${\cal
E}_0^+=\{e_{ij}\}_{i<j\in \N}$ (resp. ${\cal
E}_0^-=\{e_{ij}\}_{i>j\in \N}$) are the positive (resp. the
negative) root vectors and
$
\{e_{i,i+1}\}_{i\in \N}
$
are the simple root vectors.

\smallskip
{\it Definition 2. The Lie superalgebra $\2gl$ is a complex
linear space with a basis $\{E_{ij}\}_{i,j\in \Z}$. The
$\Z_2$-grading on $\2gl$  is defined from the requirement that
$E_{0j}$, $E_{j0}$, $0\ne j\in \Z$ are odd generators, whereas all other
generators are even. The supercommutator on  $\2gl$
is a linear extension of the relations: }

$$
\[E_{ij},E_{kl}\]=\delta _{jk}E_{il}-
(-1)^{deg(E_{ij})deg(E_{kl})}
\delta
_{il}E_{kj},\;\; i,j,k,l\in \Z. \eqno (2)
$$

\smallskip
As a basis in the Cartan subalgebra ${\cal H}$ we choose
$\{E_{ii}\}_{i\in \Z}$ with a natural order between the generators:
$E_{ii}<E_{jj}$, if $i<j$. ${\cal E}^+=\{E_{ij}\}_{i<j\in \Z}$
(resp. ${\cal E}^-=\{E_{ij}\}_{i>j\in \Z}$) are the positive (resp. the
negative) root vectors in $\2gl$ and $\{E_{i,i+1}\}_{i\in \Z}$ are the
simple root vectors.

Both algebras are isomorphic. In order to see this let
$g:\Z \rightarrow \N$ be a bijective map, defined as
$$
g(z)=2|z|+\theta (z)\in \N, \q \forall z\in \Z . \e (3)
$$
Then it is easy to verify that the map $\varphi$, which is a
linear extension of the relations
$$
\v (E_{ij})=e_{g(i),g(j)}, \;\; i,j\in \Z, \e (4)
$$
is an isomorphism of $\2gl$ on $\1gl$. Therefore both $\1gl$ and
$\2gl$ are  two different realizations of one and the same
algebra, namely $\gl$. Note that $\v$ is a map of ${\cal H}$
onto ${\cal H}_0$; it is not however a map of ${\cal E}^+$ into
${\cal E}_0^+$.  For instance take $E_{-1,0}\in {\cal E}^+$. Then
$\v(E_{-1,0})=e_{21}\in {\cal E}_0^-$.  Hence a highest
weight representation of $\2gl$ may be not a highest weight
representation of $\1gl$.

The reasons for studying representations of this particular
superalgebra, namely $\gl$, stem from physical considerations.
Our motivation originates from an attempt to introduce new
quantum statistics both in quantum mechanics$^{1, 2}$ (in this
case the superalgebras are finite-dimensional) and in
quantum field theory (QFT).$^{3,4}$ In order to see where the
connection to the statistics comes from, we recall shortly the
origin of the Lie superstatistics.

The starting point is based on the observation that any $n$ pairs
$b_1^\pm,\l,b_n^\pm$ of Bose creation and annihilation operators
(CAOs), namely  (below and throughout $[x,y]=xy-yx$, $\{x,y\}=xy+yx$)
$$
[b_i^-,b_j^+]=\delta_{ij},\quad [b_i^-,b_j^-]=[b_i^+,b_j^+]=0,\e(5)
$$
considered as odd elements, generate a representation, the Bose
representation $\rho$, of the Lie superalgebra
$osp(1|2n)\equiv B(0|n).^5$ Denote by
$B_1^\pm,\l,B_n^\pm$ those generators of $B(0|n)$, which in the
Bose representation coincide with the Bose operators,
$\rho(B_i^\pm)=b_i^\pm$. Similarly as the
Chevalley generators do, the operators $B_1^\pm,\l,B_n^\pm$ and
the relations they satisfy, namely
$$
[\{B_i^\xi,B_j^\eta\},B_k^\epsilon]=(\epsilon-\xi)\delta_{ik}B_j^\eta
+(\epsilon-\eta)\delta_{jk}B_i^\xi, \q
\xi,\eta , \epsilon =\pm \; or \; \pm1, \eqno(6)
$$
define uniquely the LS $B(0|n)$.$^{5}$
The operators $B_i^\pm$ are odd root vectors of
$B(0|n)$, whereas
$\{B_j^+,B_j^-\}$ belong to the Cartan subalgebra.
The operators (6) are known
in quantum field theory: these are the para-Bose operators,
generalizing the statistics of the tensor fields.$^{6}$  The
important conclusion is that the representation theory of $n$
pairs of para-Bose (pB) operators is equivalent to the
representation theory of the Lie superalgebra $B(0|n)$.
Certainly in QFT the algebra is $B(0|\infty)$, it is
infinite-dimensional.

The identification of the para-Bose statistics with a well known
algebraic structure provides a natural background for further
generalizations. In QFT the commutation relations between the CAOs are
determined from the translation invariance of the field under
consideration.$^{7}$ In momentum space the translation invariance of a
scalar (or tensor) field $\Psi(x)$ is expressed as a commutator
between the energy-momentum $P^m, \; m=0,1,2,3$ and the CAOs
$a_i^{\pm}$ of $\Psi(x)$: $$ [P^m, a_i^\pm]=\pm k_i^m a_i^\pm ,\e(7)
$$ where the index $i$ replaces all (continuous and discrete) indices
of the field and $$ P^m={1\over 2}\sum_j k_j^m
\{a_j^+,a_j^-\}.   \e(8) $$ To quantize the field means,
loosely speaking, to find solutions of Eqs. (7) and (8), where the
unknowns are the CAOs $a_i^\pm$.
The Bose operators (5) and their generalization, the pB operators (6),
certainly satisfy (7).  By no means however they do not exhaust the
set of the possible solutions.

The first possibility for finding new solutions and hence
for further generalization of the statistics stems from
the observation that the commutation relations between the Cartan
elements and the root vectors, in particular Eq. (7), remain unaltered
upon q-deformations. The deformations of the parastatistics
along this line was studied in Refs. 8-11 and more generally in Ref. 12.

Another opportunity,
closely related to the present paper,
is based on the observation that $B(0|n)$ belongs to the class {\it B}
superalgebras in the classification of Kac.$^{13}$ Therefore
it is natural to try to satisfy the quantization equations (7)
and (8) with CAOs, generating superalgebras from the classes {\it
A}, {\it C} and {\it D} or generating other superalgebras from the
class {\it B}. In Refs. 3, 4  it was shown that this is possible
indeed.  For charged tensor fields the main quantization
condition (7) can be satisfied with CAOs, which generate the LS
$\2gl$, namely a LS from the class {\it A}.  Up to now however
this new statistics, the $A-$superstatistics, did not achieve any
further development. The reason is that so far the Fock spaces
corresponding to the $A-$superstatistics were not constructed.
Here we come to the relation between the $A-$superstatistics and
the present investigation. The Fock spaces are representation
spaces of $\gl$. In order to study the physical consequences of
the $A-$superstatistics in QFT one has to develop first the
representation theory of $\2gl$ (for charged scalar fields) and
of $\1gl$ (for neutral fields). This is what we do in the
present paper. The reason to study only highest weight
representations reflects the fact that there should exist a state
with a lowest energy, a vacuum, which turns to be the highest
weight vector in the corresponding $\gl-$module.

So far the $A-$superstatistics was tested only in
finite-dimensional cases, namely in the frame of a (noncanonical)
quantum mechanics. We have in mind the Wigner quantum systems,
introduced in Refs. 1 and 2, which attracted recently some
attention from different points of view.$^{14,15,16}$  These
systems possess quite unconventional physical features,
properties which cannot be achieved in the frame of the canonical
quantum mechanics. The $(n+1)-$particle WQS, based on
$sl(1/3n)$,$^{17}$ exhibits a quark like structure: the composite
system occupies a small volume around the centre of mass and
within it the geometry is noncommutative.  The underlying
statistics is a Haldane exclusion statistics,$^{18}$ a subject
of considerable interest in condensed matter physics.  The
$osp(3/2)$ WQS, studied in Ref. %%%14, 
19, leads to a picture where two
spinless point particles, curling around each other, produce an
orbital (internal angular) momentum $1/2$. One can expect that
also in QFT the Lie superstatistics could lead to new  features.

In the literature one does not find many papers dealing with
representations of infinite-dimensional simple Lie
superalgebras.$^{20, 21}$ Implicitly however such algebras and
their representations were used in theoretical physics since the
QFT was created.  On the first place we have in mind  the
ordinary Fock space $W_1$ of infinitely many pairs of Bose CAOs
$\{b_i^\pm\}_{i\in \Z}$. As mentioned above, the Bose operators
are (representatives of) the odd generators of $B(0|\infty)$ and
their Fock space $W_1$ is one particular irreducible
$B(0|\infty)-$module. The Fock spaces $W_p$ of para-Bose
operators $\{B_i^\pm\}_{i\in \Z}$, corresponding to order  of the
parastatistics $p\in \N$,$^6$ are also irreducible and inequivalent to
each other $B(0|\infty)-$modules.  The Clifford construction in
Ref. 21 is a generalization to the case when both bosons
$\{b_i^\pm\}_{i\in \Z}$,
considered as odd variables, and fermions $\{f_i^\pm \}_{i\in
\Z}$, which are even generators, are involved.  The assumption is
that the bosons anticommute with the fermions.  Then any $n$
pairs of Bose CAOs and $m$ pairs of Fermi CAOs generate (a
representation of) the Lie superalgebra $B(m|n).^{22}$ Therefore
the Fock representation of $\{b_i^\pm,\;f_i^\pm \}_{i\in \Z}$ is
an irreducible $B(\infty|\infty)-$module.
Its restriction to
$gl(\infty|\infty)$ leads to a set of irreducible representations
of this superalgebra.

In the paper we use essentially results from the
representation theory of $gl(1|n)$. The finite-dimensional
irreducible modules (fidirmods) of the latter are, one can say, well
understood. A character formula for all typical$^{13}$  and
atypical$^{23}$ modules has been constructed. The dimensions of all
fidirmods are known.$^{24,25}$ A basis, similar to the GZ basis
of $gl(n)$, was defined and its transformation under the action
of the Chevalley generators was written down.$^{26,27}$ The results
were even generalized to the quantum algebra $U_q[gl(1|n)]$.$^{28}$
This is in contrast to the more general case of $gl(m|n)$
and $U_q[gl(m/n)]$,
where most of the above problems are still waiting to be solved
although partial results do exist.$^{29,30,31,32,33}$

The irreducible highest weight representations  of $\1gl$, which
we consider, are a generalization to the infinite-dimensional
case of the finite-dimensional essentially typical
representations of $gl(1|n)$ in the Gel'fand-Zetlin basis (GZ
basis). In order to see where the possibility for a
generalization comes from we recall (Sect. II.A) the way the
Gel'fand-Zetlin basis was introduced.$^{31}$  This basis is,
however, inappropriate for a generalization to the case of
highest weight $\2gl$ modules.  Therefore in Sect. II.B we modify
it, introducing a new basis, which we call a $C-$basis. It is an
analogue of the $C-$basis for $gl_\infty$.$^{34, 35}$ Section III
is devoted to the irreducible $\gl$ modules. In Sect. III.A we
extend the Gel'fand-Zetlin basis to the infinite-dimensional case
and apply it to $\1gl $.  The highest weight irreducible $\2gl$
representations are defined in Sect.  III.B. They appear as a
generalization of the essentially typical representations of
$gl(1|n)$ in the $C-$basis. The transformations of the basis
under the action of the algebra generators are explicitly written
down.

\bigskip

Throughout the paper we use the notation:

\smallskip
LS, LS's - Lie superalgebra, Lie superalgebras;

CAOs - creation and annihilation operators;

fidirmod(s) - finite-dimensional irreducible module(s);

GZ basis - Gel'fand-Zetlin basis;

$\N$ - all positive integers;

$\Z$ - all integers;

$\Z_+$ - all non-negative integers;

$\Z_2=\{\bar{0},\bar{1}\}$ - the ring of all integers modulo 2;

$\C$ - all complex numbers;

$[p;q]=\{p,p+1,p+2,\l,q-1,q\}$, if $q-p\in \Z_+ $
and $[p;q]=\emptyset$ otherwise;
\hfill (9)

$[m]_{k}=[m_{1k}, m_{2k}, \l m_{kk}],$ where $m_{ik}\in \C; \hfill (10)$

$[M]_{2k+\t }=[M_{-k,2k+\t }, M_{-k+1,2k+\t },
\l , M_{k-1+\t ,2k+\t }], \;\; \t\in \{ 0,1\},\;\;k\in \N; \hfill (11)$

$l_{1j}=m_{1j}+1,\q l_{ij}=-m_{ij}+i-1,\q i\in [2;j] ;\hfill (12)$

\smallskip
$ L_{0,2k+\t}=M_{0,2k+\t},\q  \t\in \{ 0,1\}, $

$ L_{i,2k+\t}=-M_{i,2k+\t}+i+1, \q \t\in \{ 0,1\},\q i\in [-k;-1], \hfill (13)$

$ L_{j,2k+\t}=-M_{j,2k+\t}+j-1, \q \t\in \{ 0,1\},\q j\in [1;k-1+\t];$
\smallskip

$[m]\equiv [m_1, m_2, \ldots , m_k, \ldots ]=
\{m_i |m_i \in \C , i\in \N \} ;\hfill (14)$

$[M]\equiv[\ldots , M_{-p}, \ldots, M_{-1}, M_0, M_1, \ldots, M_{q},\l ]
=\{M_i |M_i \in \C , i\in \Z \};
\hfill (15)$

\smallskip
$
P(j,l)=\cases {\hskip 0.2cm 1 & for $j\geq l$\cr  -1 & for
$j<l$\cr};  \hfill (16)
$

\smallskip
$
Q(j,l)=\cases {\hskip 0.2cm 1 & for $j> l$\cr -1 & for
$j\leq l$\cr};  \hfill (17)
$

\smallskip

$
\theta (i) = \cases
{ 1, & for $i\ge 0$ \cr
  0, &  for $i<0$. \cr}\hfill (18)
$

\vskip 1cm
%\vfill\eject
\n
{\bf II. FINITE-DIMENSIONAL ESSENTIALLY TYPICAL REPRESENTATIONS
OF $gl(1|2n)$}
\smallskip

As in the case of $\gl$ it is convenient to use two different
notation for the finite-dimensional superalgebras from this
class. In the first notation $gl_0(1|N)$ is the same as in
Definition 1, but the indices $i,j$ run from $1$ to $N+1$.
%%%%%%%%%%%%%%%%%%%%%%%%%%%%%%%%%%%%%%%%%%%%%%%%%%%%%%%%%%%%%

Then $e_{11},e_{22},\l,e_{N+1,N+1}$ is a  basis
in the Cartan subalgebra ${\cal H}_0$.
Denote by $\ep^1,\l,\ep^{N+1}$ the dual basis,
$\ep^i(e_{jj})=\delta_j^i$.
The correspondence root vector $\leftrightarrow$ root reads:
$e_{ij} \leftrightarrow \ep^i-\ep^j ,\;i\ne j=1,\ldots,N+1$;
$\Delta^0=\{\ep^i-\ep^j\}_{i\ne j \in [1;N+1]}$ is the root system;
$
\Delta^0_+=\{\ep^i-\ep^j\}_{i< j \in [1;N+1]}$ 
and
$$
\pi^0=\{\ep^1-\ep^2,\ep^2-\ep^3,\ldots,\ep^N-\ep^{N+1}\} \e(19)
$$
are the standard systems of positive roots and simple roots,
respectively. The special linear
superalgebra $sl_0(1|N)$ is a subalgebra of $gl_0(1|N)$ spanned by
all $gl_0(1|N)$ root vectors and the Cartan elements $e_{11}+e_{ii}$
for all $i\ne 1$.

Similarly, $gl(M|1|N)$ is the same as in Definition 2, but
$i,j=-M,-M+1,\l,N$ and $M,N\in \Z_+$. In particular
$\{E_{ii}\}_{i\in [-M;N]}$
is a basis in the Cartan subalgebra ${\cal H}$ with 
$\{{\cal E} ^i\}_{i\in [-M;N]}$ its dual. The simple
root vectors are $\{E_{i,i+1}\}_{i\in [-M;N-1]}$. Hence
$$
\pi=\{{\cal E} ^{-M}-{\cal E} ^{-M+1},{\cal E} ^{-M+1}-{\cal E} ^{-M+2},
\l,{\cal E} ^{-1}-{\cal E} ^0,
{\cal E} ^0-{\cal E} ^1,\l,
{\cal E} ^{N-1}-{\cal E} ^N\} \e(20)
$$
is the system of simple roots.

We have written explicitly the systems (19) and (20) in order
to underline that they contain different number of odd roots:
$\pi^0$ has only one,  $\ep^1-\ep^2$, whereas
the odd roots in $\pi$ are ${\cal E}^{-1}-{\cal E}^0,{\cal E}^0-{\cal E}^1 $.  Therefore
the systems of the simple roots of $sl_0(1|2n)$ and $sl(n|1|n)$ are
different, despite of the fact that these algebras are
isomorphic. This property demonstrates one of the essential
differences between the Lie algebras and the Lie superalgebras.
For each simple Lie algebra there exists (up to a transformation
from the Weyl group) only one system of simple roots. This is not
the case for the basic Lie superalgebras, where several
inequivalent simple root systems can be in general defined (for
more details see Ref. 36, 37, 38).  As a result one and the same
irreducible $gl(1|2n)$ module can be described with different
signatures. We shall have to take this into account in the
definition of the $C-$basis.

\bigskip
\n
{\bf A. GZ basis$^{31}$}
\smallskip

Let $V([m]_{N+1})$ be a highest weight finite-dimensional
irreducible $gl_0(1|N)$ module (fidirmod) with a highest weight
$$
[m]_{N+1}\equiv [m_{1,N+1}, m_{2,N+1}, \l , m_{N+1,N+1}]
\equiv \sum_{i=1}^{N+1} m_{i,N+1}\ep^i,\e(21)
$$
where
$$
m_{j,N+1}\in \C, \;j=1,\l,N+1, \q
m_{i,N+1}-m_{i+1,N+1}\in \Z _+, \;i=2,3,\l ,N. \e(22)
$$
If $x_{N+1}$ is the highest weight
vector in $V([m]_{N+1})$, then
$e_{ii}x_{N+1}=m_{i,N+1}x_{N+1}$.

Consider the chain of subalgebras
$$
gl_0(1|N)\supset gl_0(1|N-1)\supset gl_0(1|N-2)
\supset \l \supset gl_0(1|2)\supset gl_0(1|1)\supset gl_0(1|0)
\equiv gl_0(1).\e(23)
$$
Then $V([m]_{N+1})$ is said to be essentially typical, if
it is completely reducible with respect to any one of the
subalgebras in the chain (23). Each essentially typical
module $V([m]_{N+1})$ carries a typical representation$^{13}$ of
the special linear superalgebra $sl_0(1|n)$, but
the inverse is in general not true.

Set
$$
l_{1,N+1}=m_{1,N+1}+1; \; l_{i,N+1}=-m_{i,N+1}+i-1,
\; i=2,3,\l , N+1. \e (24)
$$

\smallskip
{\it Proposition 1.}$^{31}$ {\it The $gl_0(1|N)$ module
$V([m]_{N+1})$ is essentially typical if and only if}
$$
l_{1,N+1}\not\in  [l_{2,N+1};l_{N+1,N+1}]. \eqno(25)
$$
Let  $V([m]_{N+1})$ be an essentially
typical $gl_0(1|N)$ module and let
$$
V([m]_{N+1})\supset V([m]_{N})\supset V([m]_{N-1})\supset \l
\supset V([m]_{k+1})\supset \l
V([m]_{2}) \supset V(m_{11}) \e (26)
$$
be a flag of $gl_0(1|k)$ fidirmods $V([m]_{k+1}), \; k=0,1,2,\l ,N,$
where
$$
[m]_{k+1}\equiv [m_{1,k+1}, m_{2,k+1},\l ,m_{k+1,k+1}]\equiv
\sum_{i=1}^{k+1}m_{i,k+1}\ep^i    \e (27)
$$
is the signature of $V([m]_{k+1})$. In the ordered basis
$$
e_{11}, e_{22}, \l , e_{k+1,k+1} \e (28)
$$
of the Cartan subalgebra of $gl_0(1|k)$, $m_{i,k+1}$ is the
eigenvalue of $e_{ii}$ on the highest weight vector $x_{k+1}\in
V([m]_{k+1}),$
$$
e_{ii}x_{k+1}=m_{i,k+1}x_{k+1}, \q i=1,\l ,k+1. \e (29)
$$
Since we consider only essentially typical modules and the fidirmods
of $gl_0(1)$ are one dimensional, the flag (26) defines a vector
$|m\rangle $ in $V([m]_{N+1})$.  It turns out this vector is uniquely
defined (up to, certainly, a multiplicative constant) by the
signatures $[m]_{N+1},$ $ [m]_{N}, \l ,$ $ [m]_2,$ $ m_{11}.$
Therefore one can set

$$|m\rangle
\equiv \left[\matrix
{ [m]_{N+1} \cr
&\cr
[m]_{N} \cr
   . \cr
   . \cr
   . \cr
  [m]_2 \cr
&\cr
 m_{11}\cr }\right]
\equiv \left[\matrix
{m_{1,N+1}& m_{2,N+1}& \ldots &m_{N,N+1} &m_{N+1,N+1}\cr
&\cr
m_{1,N} & m_{2,N} &\ldots & m_{N,N} & \cr
&\cr
\ldots &\ldots &\ldots & & \cr
& \cr
m_{12} & m_{22} &&& \cr
&\cr
m_{11} & & &&\cr
}\right]. \eqno(30)
$$
The vectors (30), corresponding to all possible flags (26),
constitute a basis $\Gamma([m]_{N+1})$
in the $gl_0(1|N)$ fidirmod $V([m]_{N+1}).$
This is the GZ basis introduced in Ref. 31 (for the more general
case of $gl(M/N)$).

\smallskip
{\it Proposition 2.}$^{31}$  {\it The GZ basis $\Gamma ([m]_{N+1})$ in
the essentially typical module $V([m]_{N+1})$ is given by all tables (30)
for which

1. the numbers $m_{i,N+1}, \; i=1,2,\l N+1$ are fixed for all tables
and satisfy (22), (24), (25).

2. $m_{1i}-m_{1,i-1}\equiv \theta _{i-1}\in \{ 0,1\},
\q i=2,3,\l ,N+1, \hfill (31)$

3. $m_{i,j+1}-m_{ij}\in \Z _+;\;\; m_{ij}-m_{i+1,j+1}\in \Z_+,
\;\; 2\leq i\leq j\leq N. \hfill (32)$}

\n
{\it The transformations of the basis $\Gamma ([m]_{N+1})$ under
$gl_0(1|N)$ are completely defined from the action of the
Chevalley generators}

$$
\eqalignno{
& e_{ii}|m\rangle =(\sum_{k=1}^i m_{ki}-\sum_{k=1}^{i-1}m_{k,i-1})
|m\rangle ,\q i=1,2,\l ,N+1, &(33) \cr
&&\cr
& e_{12}|m\rangle =\theta_1|m\rangle _{(11)},\q
e_{21}|m\rangle =(1-\theta_1)(l_{12}-l_{22})|m\rangle _{-(1,1)}, & (34) \cr
&&\cr
& e_{i,i+1}|m\rangle=\theta _i(1-\theta _{i-1})|m\rangle_{(1i)}+
\sum_{j=2}^{i} \left(-
{\prod_{k=2}^{i-1} (l_{k,i-1}-l_{ji}+1)\prod_{k=2}^{i+1}
(l_{k,i+1}-l_{ji})
\over {\prod_{k\neq j=2}^{i}(l_{ki}-l_{ji})(l_{ki}-l_{ji}+1)}}
\right)^{1/2} & \cr
&& \cr
& \hskip 4.8cm \times {(l_{1i}-l_{ji})(l_{1i}-l_{ji}+1)\over
{(l_{1,i+1}-l_{ji})(l_{1,i-1}-l_{ji}+1)}}|m\rangle_{(ji)} ,
\q i=2,\l ,N, &(35)  \cr
&& \cr
& e_{i+1,i}|m\rangle=\theta _{i-1}(1-\theta _{i})
{\prod_{k=2}^{i-1}(l_{1,i+1}-l_{k,i-1}-1)\prod _{k=2}^{i+1}
(l_{1,i+1}-l_{k,i+1})\over
{\prod _{k=2}^i(l_{1,i+1}-l_{ki}-1)(l_{1,i+1}-l_{ki})}}|m\rangle_{-(1,i)}  & \cr
& +
\sum_{j=2}^{i} \left(-
{\prod_{k=2}^{i-1} (l_{k,i-1}-l_{ji})\prod_{k=2}^{i+1}
(l_{k,i+1}-l_{ji}-1)
\over {\prod_{k\neq j=2}^{i}(l_{ki}-l_{ji}-1)(l_{ki}-l_{ji})}}
\right)^{1/2}
|m\rangle_{-(ji)} , \q i=2,\l ,N, &(36)  \cr
&& \cr
\cr
}
$$
{\it where $l_{1j}=m_{1j}+1; \q l_{ij}=-m_{ij}+i-1, \q i\ne 1$ and
the table $|m\rangle_{\pm (i,j)} $ is obtained from the table
$|m\rangle $ by the
replacement $m_{ij}\rightarrow m_{ij}\pm 1.$ }

If a vector from the r.h.s. of (35) or (36) does not belong
to the module under consideration, then the corresponding term is
zero even if the coefficient in front is undefined; if an equal
number of factors in numerator and denominator are simultaneously
equal to zero, they should be canceled out.

The $gl_0(1|N)$ highest weight vector $x_{N+1}$ in $V([m]_{N+1})$
is a vector from the GZ basis
$$
x_{N+1}=|\hat{m}\rangle ,  \;\; for \;\; which \;\; m_{ii}=m_{i,i+1}=
\l =m_{i,N+1}, \q i=1,2,\l , N,\e(37)
$$
i.e.,
$$|{\hat m}\rangle
= \left[\matrix
{m_{1,N+1} & m_{2,N+1} & \ldots &m_{N,N+1} &m_{N+1,N+1}\cr
&\cr
m_{1,N+1} & m_{2,N+1} &\ldots & m_{N,N+1} & \cr
&\cr
\ldots &\ldots &\ldots & & \cr
& \cr
m_{1,N+1} & m_{2,N+1} &&& \cr
&\cr
m_{1,N+1} & & &&\cr
}\right]. \eqno(38)
$$

In this case
$$
e_{ii}|\hat{m}\rangle =m_{i,N+1}|\hat{m}\rangle , \q i=1,2,\l ,N+1,
\q e_{k,k+1}|\hat{m}\rangle =0, \q k=1,2,\l ,N. \eqno(39)
$$

%\vskip 1cm
\vfill\eject
\n
{\bf B. C-basis}

\smallskip
Let $E_{ij},\;\; i,j=-n,-n+1,\l ,n$ be the generators of
$gl(n|1|n).$ Define a sequence of subalgebras
$$
gl(k|1|k-1+\theta )=lin.env. \{ E_{ij} |\; i,j\in
[-k;k-1+\theta ]\}\q
\forall   \theta \in \{0,1\}, \;\;     k\in [1-\t ; n]. \e (40)
$$
As an ordered basis in the Cartan subalgebra of
$gl(k|1|k-1+\theta)$ take
$$
E_{-k,-k}, E_{-k+1,-k+1},\l , E_{k-1+\theta ,k-1+\theta }. \e (41)
$$

\smallskip
{\it Proposition 3.} {\it The map $\varphi ,$ which is a linear extension of
the relations
$$
\v (E_{ij})=e_{g(i),g(j)}, \;\; i,j=-n,-n+1,\l , n, \e (42)
$$
is an isomorphism of $gl(n|1|n)$ on $gl_0(1|2n)$. Its restriction on
$gl(k|1|k-1+\t )$ is an isomorphism of $gl(k|1|k-1+\t )$
on $gl_0(1|2k-1+\t )$
for each $\t \in \{0,1\}$ and $k\in [1-\t ;n].$ The chain of subalgebras
$$
gl(n|1|n)\supset gl(n|1|n-1)\supset gl(n-1|1|n-1)\supset gl(n-1|1|n-2)
\supset \l \supset gl(1|1|1)\supset gl(1|1)\supset gl(1), \e (43)
$$
$(gl(1|1|0)\equiv gl(1|1), \;\;gl(0|1|0)\equiv gl(1))$
is transformed by $\v $ into the chain (23)}
$$
gl_0(1|2n)\supset gl_0(1|2n-1)\supset gl_0(1|2n-2)
\supset \l \supset gl_0(1|2)\supset gl_0(1|1)\supset gl_0(1). \e (44)
$$
The proof is straightforward.

The isomorphism $\v $ allows one to turn any
$gl_0(1|2k-1+\t )$ irreducible module
$V([m]_{2k+\t })$ into a $gl(k|1|k-1+\t)$ module:
$$
\varphi (E_{ij})x =e_{g(i),g(j)}x , \quad \forall
x \in V([m]_{2k+\t}). \eqno(45)
$$

The relevant for us point is that each $V([m]_{2k+\t})$ can be
labeled also with its highest weight
%$[M]_{2k+\t }$
with respect
to $gl(k|1|k-1+\t )$. By definition it consists of
the eigenvalues of the representatives of the Cartan generators
(41), namely
$$
\eqalign{
&  \v (E_{-k,-k}), \v (E_{-k+1,-k+1}),\l ,
\v (E_{-2,-2}), \v (E_{-1,-1}),\v (E_{0,0}),\v (E_{1,1}),\l,
  \v (E_{k-1+\theta ,k-1+\theta })  \cr
}\e (46)
$$
on the $gl(k|1|k-1+\t )$ highest weight vector
$y_{2k+\t }\in V([m]_{2k+\t })$. The latter is defined from the
requirements
$$
\eqalignno{
& \v (E_{ij})y_{2k+\t}=0, \q i<j=-k,-k+1,
\l ,k-1+\t , &  (47)\cr
& \v (E_{ii})y_{2k+\t }=M_{i,2k+\t }y_{2k+\t }, \q i=-k,-k+1,
\l ,k-1+\t . &  (48)\cr
}
$$
Set
$$
[M]_{2k+\t }\equiv [M_{-k,2k+\t }, M_{-k+1,2k+\t },
\l , M_{k-1+\t ,2k+\t }]. \e (49)
$$
The new signature $[M]_{2k+\t }$ defines, as mentioned above,
uniquely $V([m]_{2k+\t})$. Hence
$$
V([m]_{2k+\t})=V([M]_{2k+\t}). \e(50)
$$

Consider now a GZ basis vector $|m\rangle$ corresponding to the flag
$$
V([m]_{2n+1})\supset V([m]_{2n})\supset V([m]_{2n-1})\supset \l
\supset V([m]_{2k+\t }) \supset \l
V([m]_{2}) \supset V(m_{11})
\q   \leftrightarrow \q |m\rangle,  \e (51)
$$
namely the vector (30) with $N=2n$.
In view of $(50)$ the same flag can be written as
$$
V([M]_{2n+1})\supset V([M]_{2n})\supset V([M]_{2n-1})\supset \l
\supset V([M]_{2k+\t }) \supset \l
\supset V([M]_{2}) \supset V(M_{11}) \e (52)
$$
and therefore the vector $|m\rangle$ is completely defined
by the signatures
$[M]_{2n+1},$ $ [M]_{2n}, \l ,$
$ [M]_2,$ $ M_{11}.$
Therefore we can write any GZ basis vector (30) also in the form
$$
|M\rangle \equiv \left[\matrix
{ M_{-n,2n+1} &M_{-n+1,2n+1} & \l  &M_{-1,2n+1}
&M_{0,2n+1} &M_{1,2n+1} &\ldots
&M_{n-1,2n+1} & M_{n,2n+1}\cr
&&\cr
M_{-n,2n} &M_{-n+1,2n} & \l  &M_{-1,2n}
&M_{0,2n} &M_{1,2n} &\ldots
&M_{n-1,2n} & \cr
&&\cr
&M_{-n+1,2n-1} & \l  &M_{-1,2n-1}
&M_{0,2n-1} &M_{1,2n-1} &\ldots
&M_{n-1,2n-1} & \cr
&\ldots&\ldots & \ldots & \ldots  &\ldots &\ldots \cr
&&\ldots & \ldots & \ldots  &\ldots &\ldots \cr
& & & M_{-1,3} & M_{03} &M_{13} \cr
&&\cr
& & & M_{-1,2} &M_{02} \cr
&&\cr
& & & & M_{01} \cr
}\right]. \e (53)
$$
Obviously (30) (with $N=2n$) and (53) are two different
labelings for one and the same vector $|m\rangle \equiv |M\rangle
$.  We call the basis, written in the notation (53), a
$C-$basis in $V([M]_{2n+1 })\equiv V([m]_{2n+1 })$  and denote it
as $\G ([M]_{2n+1})$.

In order to use effectively the basis $\G ([M]_{2n+1})$ we need
to determine all signatures $[M]_{2k+\t}$, namely to find the
values of the entries in (53). To this end we have to determine
as a first step the highest weight vector $y_{2k+\t }$ within
each $gl(k|1|k-1+\t)-$module  $V([m]_{2k+\t })$ in the chain
(51) and subsequently, using (48), to compute  its
$gl(k|1|k-1+\t)$ signature $[M]_{2k+\t}$.

\smallskip
{\it Proposition 4. The $gl(k|1|k-1+\t)$ highest weight vector
$y_{2k+\t}$ in $V([m]_{2k+\t})$ (from the chain (51)) is the GZ
vector $|m\rangle_{2k+\t}$, for which}
$$
\eqalignno{
& m_{1,2r+\p}+k-r=m_{1,2k+\t},\q \forall \;
\p\in\{0,1\},\;\; r\in [1-\p;k-\p];& (54) \cr
%&&\cr
& m_{r-j,2k-2j+\p}=m_{r,2k+\t},\q \forall \;
r\in [3-\t;k+1],\;\;\p\in\{0,1\},\;\;j\in[1-\t;r-2];
&(55a) \cr
%&&\cr
& m_{r-j,2k-2j+\p}=m_{r,2k+\t},\q \forall \;
r\in [k+2;2k],\;\;\p\in\{0,1\},\;\;j\in[1-\t;2k-r+\p].
&(55b) \cr
}
$$

{\it Proof:} It is easy to verify that the conditions (54) are
equivalent to
$$
\eqalignno{
& \t_{2i-1}=1, \q i\in [1;k],& (56a)\cr
&  \t_{2i}= 0, \q i\in [1;k-1+\t],& (56b)\cr
}
$$
whereas the conditions (55) can be replaced by
$$
\eqalignno{
& l_{s,2i+1}-l_{s,2i}=0, \q i\in [1;k-1+\t],\;\; s\in [2;2i],& (57a)\cr
& l_{s+1,2i}-l_{s,2i-1}-1=0, \q i\in [2;k],\;\; s\in [2;2i-1].& (57b)\cr
}
$$

We need to show that (47) holds for
$y_{2k+\t}=|m\rangle_{2k+\t}$. It certainly suffices to verify
it only for the $gl(k|1|k-1+\t)$ simple root vectors, namely to
prove that
$$
\eqalignno{
& \v(E_{-i,-i+1})|m\rangle_{2k+\t}=0, \q i\in [1;k], & (58)\cr
& \v(E_{i,i+1})|m\rangle_{2k+\t}=0, \q i\in [0;k-2+\t]. & (59)\cr
}
$$

The validity of the latter follows from the observation that 
$\; \v(E_{-1,0})=e_{21}$, $\; \v(E_{01})=[e_{12},e_{23}]$, 
$\; \v(E_{-i,-i+1})=[e_{2i,2i-1},e_{2i-1,2i-2}],\; i\in [2;k]$,
$ \; \v(E_{i-1,i})=[e_{2i-1,2i},e_{2i,2i+1}], \; i\in [2;k-1+\t]$
and Eqs. (34)-(36). This completes the proof.

We are now ready to determine the $gl(k|1|k-1+\t )$ signature
of $V([m]_{2k+\t})$ for any $\t\in \{0,1\}$ and $k\in [1;n]$.
Taking into account (54), (55) and (45) and
using the transformation relation (33), one obtains
$$
\eqalignno{
& \v(E_{ii})|m\rangle_{2k+\t}
   =e_{2|i|,2|i|}|m\rangle_{2k+\t}
   =(m_{i+k+2,2k+\t}+1)|m\rangle_{2k+\t},
  \q i\in [-k;-1] ,& (60a) \cr
& \v(E_{00})|m\rangle_{2k+\t}
   =e_{11}|m\rangle_{2k+\t}
   =(m_{1,2k+\t}-k)|m\rangle_{2k+\t},& (60b)\cr
& \v(E_{ii})|m\rangle_{2k+\t}
   =e_{2i+1,2i+1}|m\rangle_{2k+\t}
   =m_{i+k+1,2k+\t}|m\rangle_{2k+\t},
  \q i\in [1;k-1+\t] ,& (60c) \cr
}
$$
Comparing (60) with the definition (48) we obtain the
$gl(k|1|k-1+\t )$ signature $[M]_{2k+\t}$ of $V([m]_{2k+\t})$:
$$
\eqalignno{
& M_{i,2k+\t}=m_{i+k+2,2k+\t}+1, \q i\in [-k;-1] ,& (61a) \cr
& M_{0,2k+\t}=m_{1,2k+\t}-k, & (61b)\cr
& M_{i,2k+\t}=m_{i+k+1,2k+\t}, \q i\in [1,k-1+\t], & (61c) \cr
& M_{01}=m_{11}.&(61d)  \cr
}
$$
We have added the evident relation (61d) for completeness,
since it is not contained in (61a-c).
The above relations hold for any $\t\in \{0,1\}$ and $k\in [1;n]$.
In particular,
$$
\eqalignno{
& M_{i,2n+1}=m_{i+n+2,2n+1}+1, \q i\in [-n;-1] ,& (62a) \cr
& M_{0,2n+1}=m_{1,2n+1}-n, & (62b)\cr
& M_{i,2n+1}=m_{i+n+1,2n+1}, \q i\in [1,n]. & (62c) \cr
}
$$
The $gl(n|1|n)$ highest weight vector
$y_{2n+1}\equiv |{\hat M}\rl $ is the one from (53), for which
$M_{i,j}=M_{i,2n+1}$ for any admissible $i$ and $j$:
$$
|{\hat M}\rangle \equiv \left[\matrix
{ M_{-n,2n+1} &M_{-n+1,2n+1} & \l  &M_{-1,2n+1}
&M_{0,2n+1} &M_{1,2n+1} &\ldots
&M_{n-1,2n+1} & M_{n,2n+1}\cr
&&\cr
M_{-n,2n+1} &M_{-n+1,2n+1} & \l  &M_{-1,2n+1}
&M_{0,2n+1} &M_{1,2n+1} &\ldots
&M_{n-1,2n+1} & \cr
&&\cr
&M_{-n+1,2n+1} & \l  &M_{-1,2n+1}
&M_{0,2n+1} &M_{1,2n+1} &\ldots
&M_{n-1,2n+1} & \cr
&\ldots &\ldots & \ldots & \ldots  &\ldots &\ldots \cr
&&\ldots & \ldots & \ldots  &\ldots &\ldots \cr
& & & M_{-1,2n+1} & M_{0,2n+1} &M_{1,2n+1} \cr
&&\cr
& & & M_{-1,2n+1} &M_{0,2n+1} \cr
&&\cr
& & & & M_{0,2n+1} \cr
}\right]. \e (63)
$$

From (31) and (32) one derives the "in-betweenness conditions",
which define completely the new basis (53).  The transformations
of the $C-$basis are most easily written in terms of the
following variables:
$$
\eqalignno{
& L_{0,2k+\t}=M_{0,2k+\t} &\cr
& L_{i,2k+\t}=-M_{i,2k+\t}+i+1, \q i\in [-k;-1], & (64)\cr
& L_{i,2k+\t}=-M_{i,2k+\t}+i-1, \q i\in [1;k-1+\t]. &\cr
}
$$
We formulate the result as a proposition.

\smallskip
{\it Proposition 5.} {\it The $2n+1-$tuple}
$
[M]_{2n+1 }= [M_{-n,2n+1 }, M_{-n+1,2n+1 },
\l , M_{n ,2n+1 }]
$
{\it is a signature of an essentially typical $gl(n|1|n)$ module
$V([M]_{2n+1})$ if and only if}
$$
\eqalignno{
& M_{i,2n+1}\in {\bf C},\; i\in [-n;n], & (65a)\cr
& M_{i,2n+1}-M_{i+1,2n+1}\in {\bf Z_+},\;
i\in [-n;-2]\cup [1;n-1], & (65b)\cr
& M_{-1,2n+1}-M_{1,2n+1}\in \N , & (65c)\cr
& M_{0,2n+1}=L_{0,2n+1}\notin [L_{-n,2n+1};L_{n,2n+1}].& (65d) \cr
}
$$

{\it The C-basis $\G ([M]_{2n+1})$ in $V([M]_{2n+1})$ consists of all
tables (53) for which the labels}
$$
M_{i,2k+\t }, \q \t \in \{0,1\},\q k\in [1-\t ;n],
\q i\in [-k; k-1+\t], \e (66)
$$
{\it take all possible values consistent with the "in-betweenness
conditions"}
$$
\eqalignno{
& M_{i,2k+1} - M_{i, 2k }\in \Z_+, \q
  k\in [1;n], \q i\in [-k;-1]\cup[1;k-1],&(67a)\cr
& M_{i,2k-1} - M_{i, 2k }\in \Z_+, \q
  k\in [2;n], \q i\in [-k+1;-1]\cup[1;k-1],&(67b)\cr
& M_{i-1,2k} - M_{i, 2k-1 }\in \Z_+, \q
  k\in [2;n], \q i\in [-k+1;-1]\cup[2;k-1],&(67c)\cr
& M_{i-1,2k} - M_{i, 2k+1 }\in \Z_+, \q
  k\in [1;n], \q i\in [-k+1;-1]\cup[2;k],&(67d) \cr
& M_{-1,2k} - M_{1, 2k-1 }\in \N, \q
  k\in [2;n],  & (67e) \cr
& M_{-1,2k} - M_{1, 2k+1 }\in \N, \q
  k\in [1;n],  & (67f) \cr
& M_{0,2k+1} - M_{0, 2k}\equiv \psi_{2k}\in \{0,1\}, \q
  k\in [1;n],  & (67g) \cr
& M_{0,2k} - M_{0, 2k-1}\equiv \psi_{2k-1}\in\{0,-1\}, \q
  k\in [1;n],  & (67h) \cr
}
$$

{\it The transformations of the C-basis under the action of the
inverse images $\v^{-1}(e_{ii}),\; \v^{-1}(e_{i,i+1})$ and
$\v^{-1}(e_{i+1,i})$ of the $gl_0(1|2n)$ Chevalley generators
follow from (33)-(36) and (61),(62). The result reads  (we write
$E_{ij}$ instead of $\v(E_{ij}))$}:

$$
\eqalignno{
& E_{ii}|M\rangle =\left(\sum_{j=-|i|}^{|i|+\theta
(i)-1}M_{j,2|i|+\theta(i)} -
\sum_{j=-|i|+1-\theta (i)}^{|i|-1}M_{j,2|i|+\theta
(i)-1}\right) |M\rangle , \quad i\in[-n;n], & (68)\cr
&&\cr
& E_{0,-1}|M\rl=(1+\psi_1)|M\rl_{(0,1)},& (69) \cr
&&\cr
& E_{-1,0}|M\rl=-\psi_1(L_{0,2}-L_{-1,2})|M\rl_{-(0,1)},& (70) \cr
&&\cr
& E_{i-1,-i}|M\rl=(1+\psi_{2i-1})(1-\psi_{2i-2})|M\rl_{(0,2i-1)}& \cr
& +\sum_{j\ne 0=-i+1}^{i-1}
\Big(-{\prod_{k\ne 0=-i+1}^{i-2}(L_{k,2i-2}-L_{j,2i-1}+1)
\prod_{k\ne 0=-i}^{i-1}(L_{k,2i}-L_{j,2i-1}+1)
\over \prod_{k\ne 0, j; k=-i+1}^{i-1}(L_{k,2i-1}-L_{j,2i-1})
                 (L_{k,2i-1}-L_{j,2i-1}+1)}\Big)^{1/2}&\cr
& \times {(L_{0,2i-1}-L_{j,2i-1})(L_{0,2i-1}-L_{j,2i-1}+1)\over
     (L_{0,2i}-L_{j,2i-1}+1)(L_{0,2i-2}-L_{j,2i-1}+1)}|M\rl_{(j,2i-1)},
	 \q i\in [2;n],    & (71)   \cr
&&\cr
& E_{-i,i}|M\rl=-\psi_{2i}\psi_{2i-1}|M\rl_{(0,2i)}& \cr
& +\sum_{j\ne 0=-i}^{i-1}
\Big(-{\prod_{k\ne 0=-i+1}^{i-1}(L_{k,2i-1}-L_{j,2i})
\prod_{k\ne 0=-i}^{i}(L_{k,2i+1}-L_{j,2i})
\over \prod_{k\ne 0,j; k=-i}^{i-1}(L_{k,2i}-L_{j,2i})
                 (L_{k,2i}-L_{j,2i}+1)}\Big)^{1/2}&\cr
& \times {(L_{0,2i}-L_{j,2i})(L_{0,2i}-L_{j,2i}+1)\over
     (L_{0,2i+1}-L_{j,2i})(L_{0,2i-1}-L_{j,2i})}|M\rl_{(j,2i)},
	 \q i\in [1;n],    & (72)   \cr
&&\cr
& E_{i,-i}|M\rl=(1+\psi_{2i-1})(1-\psi_{2i})&\cr
&\times {\prod_{k\ne 0=-i+1}^{i-1}(L_{0,2i+1}-L_{k,2i-1})
\prod_{k\ne 0=-i}^{i}(L_{0,2i+1}-L_{k,2i+1})
\over \prod_{k\ne 0=-i}^{i-1}(L_{0,2i+1}-L_{k,2i}-1)
                 (L_{0,2i+1}-L_{k,2i})}
|M\rl_{-(0,2i)}& \cr
& +\sum_{j\ne 0=-i}^{i-1}
\Big(-{\prod_{k\ne 0=-i+1}^{i-1}(L_{k,2i-1}-L_{j,2i}-1)
\prod_{k\ne 0=-i}^{i}(L_{k,2i+1}-L_{j,2i}-1)
\over \prod_{k\ne 0,j; k=-i}^{i-1}(L_{k,2i}-L_{j,2i}-1)
                 (L_{k,2i}-L_{j,2i})}\Big)^{1/2}|M\rl_{-(j,2i)},&\cr
&\hskip 12cm	 \q i\in [1;n],    & (73)   \cr
&&\cr
& E_{-i,i-1}|M\rl=-\psi_{2i-2}\psi_{2i-1}
{\prod_{k\ne 0=-i+1}^{i-2}(L_{0,2i}-L_{k,2i-2})
\prod_{k\ne 0=-i}^{i-1}(L_{0,2i}-L_{k,2i})
\over \prod_{k\ne 0=-i+1}^{i-1}(L_{0,2i}-L_{k,2i-1})
                 (L_{0,2i}-L_{k,2i-1}+1)}
|M\rl_{-(0,2i-1)}& \cr
& +\sum_{j\ne 0=-i+1}^{i-1}
\Big(-{\prod_{k\ne 0=-i+1}^{i-2}(L_{k,2i-2}-L_{j,2i-1})
\prod_{k\ne 0=-i}^{i-1}(L_{k,2i}-L_{j,2i-1})
\over \prod_{k\ne 0,j; k=-i+1}^{i-1}(L_{k,2i-1}-L_{j,2i-1}-1)
                 (L_{k,2i-1}-L_{j,2i-1})}\Big)^{1/2}|M\rl_{-(j,2i-1)},&\cr
&\hskip 12cm	 \q i\in [2;n],    & (74)   \cr
}
$$
We have written the transformation relations of the $C-$basis
under the action of generators, which are different from the
$gl(n|1|n)$ Chevalley elements. These generators however define
completely all other generators. In this sense Eqs. (68)-(74)
are complete. We shall use them in order to derive the
transformations of the $\2gl$ irreducible modules under the
action of the Chevalley generators.

{\it Remark}. We are thankful to the referee for pointing out
that {\it Proposition 4} can be proved also without using the
transformation relations (34)-(36). To this end note 
(see Eqs. (58)-(59)) that the
$gl(k|1|k-1+\t )$ highest weight vector
$y_{2k+\t }\equiv |m\rl _{2k+\t}  \in V([m]_{2k+\t })$ 
is determined from the requirement to be annihilated
by the generators
$\{\v(E_{-i,-i+1})|\; i\in [1;k]\}\cup
 \{\v(E_{i,i+1})|\; i\in [0;k-2+\t]\}$, i.e., by
$
\{e_{2k,2k-2},e_{2k-2,2k-4},\l,e_{42},e_{21},e_{13},e_{35},\l,
e_{2k+2\t-3,2k+2\t-1}\}.
$
The roots, corresponding to the above
root vectors, namely
$$ 
\hat{\pi}_{2k+\t}=\{\ep^{2k}-\ep^{2k-2},\ep^{2k-2}-\ep^{2k-4},\l,
\ep^{4}-\ep^{2},\ep^{2}-\ep^{1},
\ep^{1}-\ep^{3},\ep^{3}-\ep^{5},\l,\ep^{2k+2\t-3}-\ep^{2k+2\t-1}\}, \e(75)
$$
can be taken as a new system of simple roots of $gl(1|2k+\t-1)$
with a system of positive roots ${\hat \Delta}
_+^{2k+\t}$.

Let $\Lambda_{2k+\t}\equiv [m]_{2k+\t}
\equiv\sum_{i=1}^{2k+\t}m_{i,2k+\t}\ep^i$ 
be the standard signature (= the highest weight) of $V([m]_{2k+\t})$, 
namely the signature corresponding to the  choice of simple roots
$$
\pi_{2k+\t}=\{\ep^1-\ep^2, \ep^2-\ep^3,\l,\ep^{2k-1+\t }-\ep^{2k+\t}  \}. 
\e(76)
$$
Denote by  $\Delta_+^{2k+\t}$ the
corresponding to it system of positive roots.
The problem is to determine the signature (= the highest weight)
${\hat \Lambda}_{2k+\t}$
of  $V([m]_{2k+\t})$ with respect to  ${\hat \Delta}_+^{2k+\t}$.
This problem can be solved on the ground of
results from Refs.$^{39,40}$  
Given a subset of positive roots
$\Delta_+'$ of $gl(1|2k+\t-1)$ 
and a simple root $\a \in \Delta_+'$, one constructs
a new system of positive roots 
${\Delta}_+''$ by a simple 
$\a$ reflection 
$\langle \a \rl$:$^{39,40}$ 
$$
{\Delta}_+''=
\langle \a \rl(\Delta_+')=
\cases {r_{\a}(\Delta_+'), & if $\; \a \; {\rm is} \;\; {\rm even;}$\cr
&\cr       
 (\Delta_+'\cup\{-\a\})\backslash \{\a\},
        & if  $\a \;{\rm is} \; {\rm odd}$  ,\cr   }\e(77)
$$
where $r_{\a}$ is an element from the Weyl group of $gl(1|2k+\t-1)$,
corresponding to $\a$.

If $V_{2k+\t}$ is an essentially typical $gl(1|2k+\t-1)$
module  with a highest weight $\lambda '$, corresponding to $\Delta_+'$, 
then the highest weight
with respect to ${\Delta}_+''$ is
$$ 
\lambda''=r_\a(\lambda')\quad {\rm if} \quad \a  \quad {\rm is} \; 
{\rm an \; even \; root \;
} \quad {\rm and} \quad
\lambda''= \lambda'-\a \quad {\rm if} \quad \a \quad {\rm is \;an \; odd \;
root}.
 \e(78) 
$$

Let $\prod_{i=1}^N \langle \a_i \rl=
\langle \a_1 \rl \langle \a_2 \rl\ldots  \langle \a_N \rl$.
Then
$$
{\hat \Delta}_+^{2k+\t}=\prod_{i=1}^k \prod_{j=1}^{2i-1}
\langle \ep_j-\ep_{2i}\rl\Delta_+^{2k+\t}. \e(79)
$$
From (77)-(79) one derives that 
$$
{\hat \Lambda}_{2k+\t}=\sum_{j=2}^{k+1}(m_{j,2k+\t}+1)\ep ^{2k-2j+4}+
(m_{1,2k+\t}-k)\ep ^1+\sum_{j=k+2}^{2k+\t}m_{j,2k+\t}\ep ^{2j-2k-1}, \e(80)
$$
i.e.,
$$
\eqalignno{
&e_{2k-2i+4,2k-2i+4}|m\rangle_{2k+\t}
   =(m_{i,2k+\t}+1)|m\rangle_{2k+\t},
  \q i\in [2;k+1] ,& (81a) \cr
&  e_{11}|m\rangle_{2k+\t}
   =(m_{1,2k+\t}-k)|m\rangle_{2k+\t},& (81b)\cr
& e_{2i-2k-1,2i-2k-1}|m\rangle_{2k+\t}
   =m_{i,2k+\t}|m\rangle_{2k+\t},
  \q i\in [k+2;2k+\t] .& (81c) \cr
}
$$
Eqs. (81) are the same as (60) (written in somewhat different notation).
Hence one obtains the $gl(k|1|k+\t-1)$ signature as given in (61)
and the corresponding to it highest weight $|m\rl_{2k+\t}$ 
({\it Proposition 4}).

\vskip 1cm
\n
{\bf III. IRREDUCIBLE REPRESENTATIONS OF $\gl$}

\bigskip
Here we construct representations of $\1gl$ and $\2gl$, which appear
as a generalization to the case $n \rightarrow \infty$ of the results
obtained in the previous section. In both cases the representations
(or the corresponding modules) are labeled with infinite sequences of
(in general different) complex numbers. Due to the isomorphism $\v$
(see (4)) each $\1gl$ module is also a $\2gl$ module and vice
versa. Therefore we can also say that we describe bellow two classes
of representations of the ``abstract'' Lie superalgebra
$gl(1|\infty)$. For definiteness we refer to the class of
representations of $\1gl$ as to Gel'fand-Zetlin (GZ) representations
(Sect. III.A), whereas the representations of $\2gl$ are said to be
C-representations.

\bigskip
\n
{\bf A. Gel'fand-Zetlin representations}

\smallskip
The extension of the results of Sect. II to the case $n\rightarrow
\infty$ is rather evident. We collect the results in a
proposition.

\smallskip
{\it Proposition 6.} {\it To each sequence of complex numbers}
$$
[m]\equiv [m_1, m_2, \ldots , m_k, \ldots ]\equiv
\{m_i |m_i \in \C , i\in \N \} ,\e (82)
$$
{\it such that }
$$
m_i-m_{i+1}\in \Z_+, \q  i=2,3,\l ,
$$%%
$$
l_{1}\not\in \{ l_{2}, l_{2}+1, l_2+2, \ldots
\}, \eqno(83)
$$
{\it where}
$$
l_{1}=m_{1}+1; \; l_{i}=-m_{i}+i-1,
\; i=2,3,\l ,  \e (84)
$$
{\it there corresponds an irreducible highest weight}
$\1gl $ module $V([m])$ {\it with a signature (82).
The basis $\Gamma ([m])$ in $V([m])$, which we call a GZ basis,
consists of all tables}
$$|m)\equiv \left[\matrix
{m_{1} & m_{2} &\ldots & m_{j} & \ldots & \l & \l  \cr
\ldots & \ldots & \ldots & \ldots & \ldots & \l \cr
\ldots & \ldots & \ldots & \ldots & \ldots  \cr
m_{1j} & m_{2j} &\ldots & m_{jj} & \cr
\ldots &\ldots &\ldots & & \cr
\ldots &\ldots &\ldots & & \cr
m_{12} & m_{22} &&& \cr
m_{11} & & &&\cr
}\right]\equiv \left[\matrix
{ [m]  \cr
 .  \cr
 .  \cr
[m]_j  \cr
% .  \cr
 .  \cr
 .  \cr
[m]_2  \cr
m_{11} \cr
}\right] ,
\eqno(85)
$$
{\it characterized by an infinite number of coordinates
$$
m_{ij}, \quad \forall  j\in \N, \quad i=1,2, \ldots , j, \eqno(86)
$$
which are consistent with the conditions:

\noindent
1. for each table $|m)$ there exists a positive $($depending
on $|m))$ integer  $N[|m)] \in \N$  such that
$$
m_{ij}=m_i, \quad \forall j>N[|m)], \quad i=1, \ldots , j;
\eqno(87)
$$
2. $m_{1i}-m_{1,i-1}\equiv \theta _{i-1}\in \{ 0,1\},
\q i=2,3,\l ;\hfill (88)$

\n
3. $m_{i,j+1}-m_{ij}\in \Z _+;\;\; m_{ij}-m_{i+1,j+1}\in \Z_+,
\;\; 2\leq i\leq j\in \N. \hfill (89)$

\n
The transformation of the basis $(85)$ is determined from
the action of the Chevalley
generators }

$$
\eqalignno{
& e_{ii}|m)=(\sum_{k=1}^i m_{ki}-\sum_{k=1}^{i-1}m_{k,i-1})|m),
\q i\in \N, &(90) \cr
&&\cr
& e_{12}|m)=\theta_1|m)_{11},\q
e_{21}|m)=(1-\theta_1)(l_{12}-l_{22})|m)_{-(1,1)}, & (91) \cr
&&\cr
& e_{i,i+1}|m)=\theta _i(1-\theta _{i-1})|m)_{(1i)}+
\sum_{j=2}^{i} \left(-
{\prod_{k=2}^{i-1} (l_{k,i-1}-l_{ji}+1)\prod_{k=2}^{i+1}
(l_{k,i+1}-l_{ji})
\over {\prod_{k\neq j=2}^{i}(l_{ki}-l_{ji})(l_{ki}-l_{ji}+1)}}
\right)^{1/2} & \cr
&& \cr
& \hskip 4.8cm \times {(l_{1i}-l_{ji})(l_{1i}-l_{ji}+1)\over
{(l_{1,i+1}-l_{ji})(l_{1,i-1}-l_{ji}+1)}}|m)_{(ji)} ,\q i=2,3, \l , &(92)  \cr
&& \cr
& e_{i+1,i}|m)=\theta _{i-1}(1-\theta _{i})
{\prod_{k=2}^{i-1}(l_{1,i+1}-l_{k,i-1}-1)\prod _{k=2}^{i+1}
(l_{1,i+1}-l_{k,i+1})\over
{\prod _{k=2}^i(l_{1,i+1}-l_{ki}-1)(l_{1,i+1}-l_{ki})}}|m)_{-(1,i)}  & \cr
& +
\sum_{j=2}^{i} \left(-
{\prod_{k=2}^{i-1} (l_{k,i-1}-l_{ji})\prod_{k=2}^{i+1}
(l_{k,i+1}-l_{ji}-1)
\over {\prod_{k\neq j=2}^{i}(l_{ki}-l_{ji}-1)(l_{ki}-l_{ji})}}
\right)^{1/2}
|m)_{-(ji)} , \q i=2,3,\l . &(93)  \cr
&& \cr
}
$$
{\it The highest weight vector $|{\hat m})$ is the one from $(85)$ for
which}
$$
m_{ij}=m_i, \q \forall j\in \N , \q i\in 1,2,\l ,j. \e (94)
$$
{\it Proof:} Let

$$|m)\equiv \left[\matrix
{[m]\cr
 .  \cr
 .  \cr
 .  \cr
[m]_{N+1}  \cr
 .  \cr
 .  \cr
 .  \cr
[m]_2  \cr
m_{11} \cr
}\right] \in \Gamma([m]). \eqno(95)$$
Then

\n
(i) $[m]_{N+1}\equiv [m_{1,N+1},m_{2,N+1}, \ldots , m_{N+1,N+1}],
\quad N=1,2,\ldots ,$
is said to be the $(N+1)^{th}-$signature of $|m);$

\n
(ii)
$$|m)^{up(N+1)}\equiv \left[\matrix
{[m]\cr
 .  \cr
 .  \cr
 .  \cr
[m]_j  \cr
 .  \cr
 .  \cr
 .  \cr
[m]_{N+2}  \cr
}\right] \;and \quad
|m)^{low(N+1)}\equiv \left[\matrix
{[m]_{N+1}\cr
 .  \cr
 .  \cr
 .  \cr
[m]_i  \cr
 .  \cr
 .  \cr
 .  \cr
[m]_2  \cr
m_{11} \cr
}\right]
\eqno(96)
$$
are said to be the $(N+1)^{th}-$upper and the $(N+1)^{th}-$lower
part of $|m),$ respectively.  Consider the subalgebra
$$
gl_0(1|N)=\{ e_{ij}|i,j=1,\ldots ,N+1\} \subset \1gl. \eqno(97)
$$
{\it Observation 1:} Let $e$ be a $gl_0(1|N)$ generator or any polynomial of
$gl_0(1|N)$ generators. Then, for any $|m)\in \Gamma ([m]), $  $e|m)$ is a
linear combination of vectors from $\Gamma ([m])$ with one and same
$(N+1)^{th}-$upper part $|m)^{up(N+1)}.$

Denote by
$$
\Gamma ([m]_i|i\geq N+1)\subset \Gamma ([m]) \eqno(98)
$$
the set of all vectors (85), that have one and the same $[m]_i$ signatures,
for all $i\geq N+1.$ Let
$$
V([m]_i|i\geq N+1) \subset V([m]) \eqno(99)
$$
be the linear span of $\Gamma ([m]_i|i\geq N+1).$ From (90)-(93) it
follows that $V([m]_i|i\geq N+1)$ is invariant with respect to $gl_0(1|N)$.
To each vector
$|m)\in \Gamma ([m]_i|i\geq N+1)$
put in correspondence
its $(N+1)^{th}-$lower part:
$$
f(|m))=|m)^{low(N+1)}, \q \forall \; |m)\in \Gamma ([m]_i|\;i\geq N+1).
\e (100)
$$
Let
$$
\Gamma ([m]_{N+1})=\{ f(|m)) \;|\; |m)\in \Gamma ([m]_i|\; i\geq N+1)\}.
\e (101)
$$
\n
Then $f$ maps bijectively  $\Gamma ([m]_{i})|i\geq N+1)$ on
$\Gamma ([m]_{N+1})$.  Obviously $\Gamma ([m]_{N+1})$ consists of
all GZ tables of an essentially typical $gl_0(1|N)$ module with a
signature $[m]_{N+1}.$ Define an action of $gl_0(1|N)$ on $|m)\in
\Gamma ([m]_{N+1})$ with the relations (33)-(36). Then the
linear envelope $V([m]_{N+1})$ of $\Gamma ([m]_{N+1})$ is an
essentially typical $gl_0(1|N)$ module with a signature
$[m]_{N+1}.$ After comparing the relations (90)-(93) with
(33)-(36) and having in mind {\it Observation 1} we have:

{\it Observation 2. } The subspace $V([m]_i|i\geq N+1) \subset V([m])$
is an essentially typical finite-dimensional $gl_0(1|N)$ module with a
signature $[m]_{N+1}$
and a GZ basis $\Gamma ([m]_i|i\geq {N+1}).$

Let $e_{ij}, e_{kl}$ be any two generators from $\1gl$ and $|m)$
be an arbitrary vector from $\Gamma([m]).$ Consider $e_{ij}, e_{kl}$ as
elements from $gl_0(1|N) \subset \1gl,$ where $N+1\geq max(i,j,k,l).$
Then $|m)$ is a vector from  the $gl_0(1|N)$ fidirmod
$V([m]_i|i\geq N+1) \subset V([m])$ and therefore ({\it Observation 2})
$$
(e_{ij}e_{kl}-(-1)^{deg(e_{ij})deg(e_{kl})}e_{kl}e_{ij})|m)=
(\delta _{jk}e_{il}-(-1)^{deg(e_{ij})deg(e_{kl})}\delta _{li}e_{kj})
|m). \eqno(102)
$$
Therefore the linear space $V([m])$ is  a $\1gl$ module.

Consider any two vectors $x, y \in V([m]),$
$$
x=\sum_{i=1}^p\alpha _i|m^i), \; y=\sum_{i=p+1}^q\alpha _i|m^i),
\;|m^i)\in \Gamma([m]), \;
$$
$$ \hskip 2cm  \alpha _i\in \C,\;i=1,\ldots ,q. \eqno(103)$$
Let
$$
\tilde{N}=max\{ N[|m^i)]|i=1,\ldots ,q \}. \eqno(104)
$$
According to (87) all vectors $|m^i), \; i=1,\ldots , q,$
have one and
the same $k-1$ signatures, for every $k-1\geq \tilde{N}.$
Therefore $|m^i) \in
V([m]_{k-1}|k-1\geq \tilde{N})\subset V([m]). $
Hence $x,y\in V([m]_{k-1}|k-1\geq \tilde{N}).$
The space $V([m]_{k-1}|k-1\geq \tilde{N})$
is a $gl_0(1|\tilde{N})$ fidirmod ({\it Observation 2}) and, therefore,
there exist a polynomial $P$ of the $gl_0(1|\tilde{N})$
generators such that $y=Px.$ Hence $V([m])$ is an irreducible
$\1gl $ module.

Consider the vector $|{\hat m})\in \Gamma ([m])$ [see (91)].
From Eqs. (90)-(93) we have
$$
e_{ii}|{\hat m})=m_i|{\hat m}), \q \forall i\in \N , \e (105)
$$
and
$$
e_{k,k+1}|{\hat m})=0, \q \forall k\in \N. \e (106)
$$
Therefore the irreducible $\1gl $ module $V([m])$ is a highest
weight module with a signature
$$
[m]\equiv [m_1, m_2, \ldots , m_k, \ldots ] \e (107)
$$
and a highest weight vector $|{\hat m})$.
This completes the proof.

\bigskip
\n
{\bf B. C-representations}

Most of the preliminary work for constructing the
representations of $\2gl$ was done in Sect. II.B.
It remains to give a precise definition of the
$C-$basis in the infinite-dimensional case and to
write down the transformation of the basis under
the action of the Chevalley generators.

Let
$$
[M]\equiv[\ldots , M_{-p}, \ldots, M_{-1}, M_0, M_1, M_2,\ldots]\equiv
\{M_i \}_{i\in \Z}    \eqno(108)
$$
be a sequence of complex numbers such that
$$
\eqalignno{
& M_{i}-M_{i+1}\in {\bf Z_+},\;
i\in [-\infty;-2]\cup [1;\infty], & (109a)\cr
& M_{-1}-M_{1}\in \N , & (109b)\cr
& M_{0}+M_1 \notin \Z. & (109c) \cr
}
$$
Here and throughout
$$
\eqalignno{
& [-\infty;a]=\{a,a-1,a-2,\l,a-i,\l\}\equiv \{a-i\}_{i\in \Z_+},&
(110)\cr
& [b;\infty]=\{b,b+1,b+2,\l,b+i,\l \}\equiv \{b+i\}_{i\in \Z_+},&
(111)\cr
}
$$
A table $|M)$, consisting of infinitely many complex numbers
$$
M_{i, 2k+\theta -1}, \; \forall k \in \N, \; \theta \in \{0,1\},
\;
i=[-k-\theta +1; k-1], \e (112)
$$
will be called a $C-$table, provided the following conditions hold:

\noindent
(1) There exists a positive, depending on $|M)$, integer
$N[|M)]$ such that
$$
 M_{i,2k+\theta -1}=M_i, \;\; \forall \; k >N[|M)], \quad
 \theta \in \{0,1\}, \;\; i \in [1-\theta -k, k-1]; \e (113)
$$
(2) The coordinates $M_{i,2k+\theta -1}$,  $\t \in \{0,1\}$,
take all possible values
$$
\eqalignno{
& M_{i,2k+1-2\t} - M_{i, 2k }\in \Z_+, \q
  k\in [1+\t;\infty], \;\; i\in [-k+\t;-1]\cup[1;k-1], & (114a) \cr
& M_{i-1,2k} - M_{i, 2k+1-2\t }\in \Z_+, \q
  k\in [1+\t;\infty], \;\; i\in [-k+1;-1]\cup[2;k-\t], & (114b) \cr
& M_{-1,2k} - M_{1, 2k+1-2\t }\in \N, \q
    k\in [1+\t;\infty],  & (114c) \cr
& M_{0,2k+1-\t} - M_{0, 2k-\t}\equiv \psi_{2k-\t}\in \{0,1-2\t\}, \q
    k\in [1;\infty].  & (114d) \cr
}
$$

Order the complex numbers  $M_{i, 2k+\theta -1}$,
$k\in \N,\;\; \t \in \{0,1\},$ as in the table below
$$
|M)\equiv \left[\matrix
{ .., & M_{1-\theta -k},& \ldots, &M_{-1},&M_0, &M_1, &\ldots,
&M_{k-1},...\cr
.., & \ldots & \ldots & \ldots & \ldots &\ldots  &\ldots &\ldots \cr
&M_{1-\theta -k,2k+\theta -1}, & \ldots, &M_{-1,2k+\theta -1},
&M_{0,2k+\theta -1}, & M_{1,2k+\theta -1}, & \ldots,
& M_{k-1,2k+\theta -1} \cr
&\ldots &\ldots & \ldots & \ldots  &\ldots &\ldots \cr
& & & M_{-1,3}, & M_{03}, &M_{13} \cr
& & & M_{-1,2}, &M_{02} \cr
& & & & M_{01} \cr
}\right], \eqno(115)
$$
We are ready now to state our main and final result.

\smallskip
{\it Proposition 7.} {\it To each sequence (108) (see also (109))
there corresponds an irreducible highest weight $\2gl$ module
$V([M])$ with a signature $[M]$.  The basis $\Gamma ([M])$ in
$V([M])$ consists of all $C-$tables (115).  The transformations of
the basis under the action of the $\2gl$ Chevalley generators
read:}

$$
\eqalignno{
& E_{kk}|M)=\left(\sum_{i=-|k|}^{|k|+\theta
(k)-1}M_{i,2|k|+\theta(k)} -
\sum_{i=-|k|+1-\theta (k)}^{|k|-1}M_{i,2|k|+\theta
(k)-1}\right) |M), \quad k\in \Z, &(116)\cr
&&\cr
& E_{0,-1}|M)=(1+\psi _1)|M)_{( 01) } & (117)\cr
&&\cr
& E_{-1,0}|M)=-\psi _1(L_{0,2}-L_{-1,2})|M)_{-( 01) }
& (118)\cr
&&\cr
&E_{01}|M)=-\psi _2(1+2\psi _1)|M)_{(0,2)}^{( 01)} &\cr
&&\cr
&+(1+\psi _1)\left( -(L_{-1,3}-L_{-1,2})
(L_{13}-L_{-1,2})\right)^{1/2}
{(L_{02}-L_{-1,2})(L_{02}-L_{-1,2}+1)
\over{(L_{03}-L_{-1,2})(L_{01}-L_{-1,2})(L_{01}-L_{-1,2}+1)}}
|M)_{( -1,2)}^{( 01)} &(119)\cr
&&  \cr
&E_{10}|M)=-(-1)^{\psi _1}(1-\psi _2)
{(L_{02}-L_{-1,2}-\psi _1-1)(L_{03}-L_{-1,3})(L_{03}-L_{13})
\over{(L_{03}-L_{-1,2}-1)(L_{03}-L_{-1,2})}}
|M)_{-( 02)}^{-( 01)} &\cr
&&\cr
&-\psi _1\left( -(L_{-1,3}-L_{-1,2}-1)
(L_{13}-L_{-1,2}-1)\right)^{1/2}
|M)_{-( -1,2)}^{-( 01)} &(120) \cr
&&\cr
&E_{k,k+1}|M)=-\psi _{2k+2}(1-\psi _{2k})(1+2\psi _{2k+1})
|M)_{(0,2k+2)}^{(0,2k+1)} & \cr
&&\cr
&+\sum_{j\ne 0=-k}^{k} \psi _{2k+2}\psi _{2k+1}
\left(-
{\prod_{i\ne 0=-k}^{k-1}(L_{i,2k}-L_{j,2k+1}+1)
\prod_{i\ne 0=-k-1}^{k}(L_{i,2k+2}-L_{j,2k+1}+1)
\over
{\prod_{i\ne 0,j;i=-k}^{k}(L_{i,2k+1}-L_{j,2k+1})
(L_{i,2k+1}-L_{j,2k+1}+1)
}}\right)^{1/2} & \cr
&&\cr
& \times
{(L_{0,2k+1}-L_{j,2k+1})
(L_{0,2k+1}-L_{j,2k+1}+1)
\over
{(L_{0,2k+2}-L_{j,2k+1}+2)
(L_{0,2k+2}-L_{j,2k+1}+1)
(L_{0,2k}-L_{j,2k+1}+1)
}}|M)_{(0,2k+2)}^{(j,2k+1)} & \cr
&&\cr
&+\sum_{j\ne 0=-k-1}^{k} (1+\psi _{2k+1})(1-\psi _{2k})
\left(-
{\prod_{i\ne 0=-k}^{k}(L_{i,2k+1}-L_{j,2k+2})
\prod_{i\ne 0=-k-1}^{k+1}(L_{i,2k+3}-L_{j,2k+2})
\over
{\prod_{i\ne 0,j; i=-k-1}^{k}(L_{i,2k+2}-L_{j,2k+2})
(L_{i,2k+2}-L_{j,2k+2}+1)
}}\right)^{1/2} & \cr
&&\cr
& \times
{(L_{0,2k+2}-L_{j,2k+2})
(L_{0,2k+2}-L_{j,2k+2}+1)
\over
{(L_{0,2k+3}-L_{j,2k+2})
(L_{0,2k+1}-L_{j,2k+2})
(L_{0,2k+1}-L_{j,2k+2}+1)
}}|M)_{(j,2k+2)}^{(0,2k+1)} &
\cr
&&\cr
&+\sum_{l\ne 0=-k-1}^{k} \; \sum_{j\ne 0=-k}^{k} Q(j,l )
 \left(-
{\prod_{i\ne 0, j;i=-k}^{k}(L_{i,2k+1}-L_{l,2k+2})
\prod_{i\ne 0=-k-1}^{k+1}(L_{i,2k+3}-L_{l,2k+2})
\over
{\prod_{i\ne 0, l; i=-k-1}^{k}(L_{i,2k+2}-L_{l,2k+2})
(L_{i,2k+2}-L_{l,2k+2}+1)
}}\right)^{1/2} & \cr
&&\cr
&\times \left(
{\prod_{i\ne 0=-k}^{k-1}(L_{i,2k}-L_{j,2k+1}+1)
\prod_{i\ne 0,l;i=-k-1}^{k}(L_{i,2k+2}-L_{j,2k+1}+1)
\over
{\prod_{i\ne 0,j; i=-k}^{k}(L_{i,2k+1}-L_{j,2k+1})
(L_{i,2k+1}-L_{j,2k+1}+1)
}}\right)^{1/2} & \cr
&&\cr
& \times
{(L_{0,2k+2}-L_{l,2k+2})
(L_{0,2k+2}-L_{l,2k+2}+1)
(L_{0,2k+1}-L_{j,2k+1})
(L_{0,2k+1}-L_{j,2k+1}+1)
\over
{(L_{0,2k+3}-L_{l,2k+2})
(L_{0,2k+1}-L_{l,2k+2})
(L_{0,2k+2}-L_{j,2k+1}+1)
(L_{0,2k}-L_{j,2k+1}+1)
}}|M)_{(l,2k+2)}^{(j,2k+1)}, & \cr
&\hskip 13cm  k\in [1,\infty], & (121) \cr
&&\cr
&E_{-k+1,-k}|M)=-(1+\psi_{2k-1}) \psi _{2k-3}(1-2\psi _{2k-2})
|M)_{(0,2k-1)}^{(0,2k-2)} & \cr
&&\cr
&-\sum_{j\ne 0=-k+1}^{k-2} (1+ \psi _{2k-1})(1-\psi _{2k-2})
\left(-
{\prod_{i\ne 0=-k+2}^{k-2}(L_{i,2k-3}-L_{j,2k-2})
\prod_{i\ne 0=-k+1}^{k-1}(L_{i,2k-1}-L_{j,2k-2})
\over
{\prod_{i\ne 0,j;i=-k+1}^{k-2}(L_{i,2k-2}-L_{j,2k-2})
(L_{i,2k-2}-L_{j,2k-2}+1)
}}\right)^{1/2} & \cr
&&\cr
& \times
{(L_{0,2k-2}-L_{j,2k-2})
(L_{0,2k-2}-L_{j,2k-2}+1)
\over
{(L_{0,2k-1}-L_{j,2k-2})
(L_{0,2k-1}-L_{j,2k-2}+1)
(L_{0,2k-3}-L_{j,2k-2})
}}|M)_{(0,2k-1)}^{(j,2k-2)} & \cr
&&\cr
&-\sum_{j\ne 0=-k+1}^{k-1} \psi _{2k-2}\psi_{2k-3}
\left(-
{\prod_{i\ne 0=-k+1}^{k-2}(L_{i,2k-2}-L_{j,2k-1}+1)
\prod_{i\ne 0=-k}^{k-1}(L_{i,2k}-L_{j,2k-1}+1)
\over
{\prod_{i\ne 0,j; i=-k+1}^{k-1}(L_{i,2k-1}-L_{j,2k-1})
(L_{i,2k-1}-L_{j,2k-1}+1)
}}\right)^{1/2} & \cr
&&\cr
& \times
{(L_{0,2k-1}-L_{j,2k-1})
(L_{0,2k-1}-L_{j,2k-1}+1)
\over
{(L_{0,2k}-L_{j,2k-1}+1)
(L_{0,2k-2}-L_{j,2k-1}+1)
(L_{0,2k-2}-L_{j,2k-1}+2)
}}|M)_{(j,2k-1)}^{(0,2k-2)} &  \cr
&&\cr
&+\sum_{l\ne 0=-k+1}^{k-1} \; \sum_{j\ne 0=-k+1}^{k-2} P(j,l )
 \left(-
{\prod_{i\ne 0,j; i=-k+1}^{k-2}(L_{i,2k-2}-L_{l,2k-1}+1)
\prod_{i\ne 0=-k}^{k-1}(L_{i,2k}-L_{l,2k-1}+1)
\over
{\prod_{i\ne 0, l; i=-k+1}^{k-1}(L_{i,2k-1}-L_{l,2k-1})
(L_{i,2k-1}-L_{l,2k-1}+1)
}}\right)^{1/2} & \cr
&&\cr
&\times \left(
{\prod_{i\ne 0=-k+2}^{k-2}(L_{i,2k-3}-L_{j,2k-2})
\prod_{i\ne 0,l; i=-k+1}^{k-1}(L_{i,2k-1}-L_{j,2k-2})
\over
{\prod_{i\ne 0,j; i=-k+1}^{k-2}(L_{i,2k-2}-L_{j,2k-2})
(L_{i,2k-2}-L_{j,2k-2}+1)
}}\right)^{1/2} & \cr
&&\cr
& \times
{(L_{0,2k-1}-L_{l,2k-1})
(L_{0,2k-1}-L_{l,2k-1}+1)
(L_{0,2k-2}-L_{j,2k-2})
(L_{0,2k-2}-L_{j,2k-2}+1)
\over
{(L_{0,2k}-L_{l,2k-1}+1)
(L_{0,2k-2}-L_{l,2k-1}+1)
(L_{0,2k-1}-L_{j,2k-2})
(L_{0,2k-3}-L_{j,2k-2})
}}|M)_{(l,2k-1)}^{(j,2k-2)}, & \cr
& \hskip 13cm k\in [2,\infty],& (122) \cr
&E_{k+1,k}|M)=-(-1)^{\psi_{2k+1}}\psi _{2k}(1-\psi _{2k+2})
&\cr
%&& \cr
& \times
{\prod_{i\ne 0=-k}^{k-1}(L_{0,2k+2}-L_{i,2k}-\psi _{2k+1}-1)
\prod_{i\ne 0=-k-1}^k(L_{0,2k+2}-L_{i,2k+2}-\psi _{2k+1}-1)
\over
{\prod_{i\ne 0=-k}^{k}(L_{0,2k+2}-L_{i,2k+1}-\psi _{2k+1}-1)
(L_{0,2k+2}-L_{i,2k+1}-\psi _{2k+1})}} &\cr
&&\cr
& \times
{\prod_{i\ne 0=-k}^{k}(L_{0,2k+3}-L_{i,2k+1})
\prod_{i\ne 0=-k-1}^{k+1}(L_{0,2k+3}-L_{i,2k+3})
\over
{\prod_{i\ne 0=-k-1}^{k}(L_{0,2k+3}-L_{i,2k+2}-1)
(L_{0,2k+3}-L_{i,2k+2})}}
|M)_{-(0,2k+2 )}^{-(0,2k+1)} & \cr
&&\cr
&-\sum_{j\ne 0=-k}^{k} (1+\psi _{2k+1})(1-\psi _{2k+2})
\left(-
{\prod_{i\ne 0=-k}^{k-1}(L_{i,2k}-L_{j,2k+1})
\prod_{i\ne 0=-k-1}^{k}(L_{i,2k+2}-L_{j,2k+1})
\over
{\prod_{i\ne 0,j;i=-k}^{k}(L_{i,2k+1}-L_{j,2k+1}-1)
(L_{i,2k+1}-L_{j,2k+1})
}}\right)^{1/2} & \cr
&&\cr
& \times
{\prod_{i\ne 0,j; i=-k}^k
(L_{0,2k+3}-L_{i,2k+1})
\prod_{i\ne 0=-k-1}^{k+1}(L_{0,2k+3}-L_{i,2k+3})
\over
{\prod_{i\ne 0=-k-1}^{k}(L_{0,2k+3}-L_{i,2k+2}-1)
(L_{0,2k+3}-L_{i,2k+2})
}}|M)_{-(0,2k+2)}^{-(j,2k+1)} & \cr
&&\cr
&-\sum_{j\ne 0=-k-1}^{k} \psi _{2k}\psi _{2k+1}
\left(-
{\prod_{i\ne 0=-k}^{k}(L_{i,2k+1}-L_{j,2k+2}-1)
\prod_{i\ne 0=-k-1}^{k+1}(L_{i,2k+3}-L_{j,2k+2}-1)
\over
{\prod_{i\ne 0,j; i=-k-1}^{k}(L_{i,2k+2}-L_{j,2k+2}-1)
(L_{i,2k+2}-L_{j,2k+2})
}}\right)^{1/2} &  \cr
&&\cr
& \times
{\prod_{i\ne 0,j; i=-k-1}^k
(L_{0,2k+2}-L_{i,2k+2})
\prod_{i\ne 0=-k}^{k-1}(L_{0,2k+2}-L_{i,2k})
\over
{\prod_{i\ne 0=-k}^{k}(L_{0,2k+2}-L_{i,2k+1})
(L_{0,2k+2}-L_{i,2k+1}+1)
}}|M)_{-(j,2k+2)}^{-(0,2k+1)} &  \cr
&&\cr
&+\sum_{l\ne 0=-k-1}^{k} \; \sum_{j\ne 0=-k}^{k} Q(j,l )
 \left(-
{\prod_{i\ne 0,j; i=-k}^{k}(L_{i,2k+1}-L_{l,2k+2}-1)
\prod_{i\ne 0=-k-1}^{k+1}(L_{i,2k+3}-L_{l,2k+2}-1)
\over
{\prod_{i\ne 0,l; i=-k-1}^{k}(L_{i,2k+2}-L_{l,2k+2}-1)
(L_{i,2k+2}-L_{l,2k+2})
}}\right)^{1/2} & \cr
&&\cr
&\times \left(
{\prod_{i\ne 0=-k}^{k-1}(L_{i,2k}-L_{j,2k+1})
\prod_{i\ne 0,l; i=-k-1}^{k}(L_{i,2k+2}-L_{j,2k+1})
\over
{\prod_{i\ne 0,j; i=-k}^{k}(L_{i,2k+1}-L_{j,2k+1}-1)
(L_{i,2k+1}-L_{j,2k+1})
}}\right)^{1/2}  |M)_{-(j,2k+1)}^{-(l,2k+2)},\q k\in [1,\infty],& (123) \cr
&& \cr
&&\cr
&E_{-k,-k+1}|M)=-(-1)^{\psi _{2k-2}}(1+\psi _{2k-3})\psi _{2k-1}&\cr
& \times
{\prod_{i\ne 0=-k+2}^{k-2}(L_{0,2k-1}-L_{i,2k-3}-\psi _{2k-2})
\prod_{i\ne 0=-k+1}^{k-1}(L_{0,2k-1}-L_{i,2k-1}-\psi _{2k-2})
\over
{\prod_{i\ne 0=-k+1}^{k-2}(L_{0,2k-1}-L_{i,2k-2}-1)
(L_{0,2k-1}-L_{i,2k-2}-2\psi _{2k-2})}} &\cr
&&\cr
& \times
{\prod_{i\ne 0=-k+1}^{k-2}(L_{0,2k}-L_{i,2k-2})
\prod_{i\ne 0=-k}^{k-1}(L_{0,2k}-L_{i,2k})
\over
{\prod_{i\ne 0=-k+1}^{k-1}(L_{0,2k}-L_{i,2k-1})
(L_{0,2k}-L_{i,2k-1}+1)}}
|M)_{-(0,2k-1)}^{-(0,2k-2)} & \cr
&&\cr
&+\sum_{j\ne 0=-k+1}^{k-2} \psi _{2k-2}\psi _{2k-1}
\left(-
{\prod_{i\ne 0=-k+2}^{k-2}(L_{i,2k-3}-L_{j,2k-2}-1)
\prod_{i\ne 0=-k+1}^{k-1}(L_{i,2k-1}-L_{j,2k-2}-1)
\over
{\prod_{i\ne 0,j; i=-k+1}^{k-2}(L_{i,2k-2}-L_{j,2k-2}-1)
(L_{i,2k-2}-L_{j,2k-2})
}}\right)^{1/2} & \cr
&&\cr
& \times
{\prod_{i\ne0,j;i=-k+1}^{k-2}
(L_{0,2k}-L_{i,2k-2})
\prod_{i\ne 0=-k}^{k-1}(L_{0,2k}-L_{i,2k})
\over
{\prod_{i\ne 0=-k+1}^{k-1}(L_{0,2k}-L_{i,2k-1})
(L_{0,2k}-L_{i,2k-1}+1)
}}|M)_{-(0,2k-1)}^{-(j,2k-2)} & \cr
&&\cr
&+\sum_{j\ne 0=-k+1}^{k-1} (1+\psi _{2k-3})(1-\psi _{2k-2})
\left(-
{\prod_{i\ne 0=-k+1}^{k-2}(L_{i,2k-2}-L_{j,2k-1})
\prod_{i\ne 0=-k}^{k-1}(L_{i,2k}-L_{j,2k-1})
\over
{\prod_{i\ne 0,j; i=-k+1}^{k-1}(L_{i,2k-1}-L_{j,2k-1}-1)
(L_{i,2k-1}-L_{j,2k-1})
}}\right)^{1/2} &  \cr
&&\cr
& \times
{\prod_{i\ne 0,j;i=-k+1}^{k-1}
(L_{0,2k-1}-L_{i,2k-1})
\prod_{i\ne 0=-k+2}^{k-2}(L_{0,2k-1}-L_{i,2k-3})
\over
{\prod_{i\ne 0=-k+1}^{k-2}(L_{0,2k-1}-L_{i,2k-2}-1)
(L_{0,2k-1}-L_{i,2k-2})
}}|M)_{-(j,2k-1)}^{-(0,2k-2)} &  \cr
&&\cr
&+\sum_{l\ne 0=-k+1}^{k-1} \; \sum_{j\ne 0=-k+1}^{k-2} P(j,l )
 \left(-
{\prod_{i\ne 0,j;i=-k+1}^{k-2}(L_{i,2k-2}-L_{l,2k-1})
\prod_{i\ne 0=-k}^{k-1}(L_{i,2k}-L_{l,2k-1})
\over
{\prod_{i\ne 0,l;i=-k+1}^{k-1}(L_{i,2k-1}-L_{l,2k-1}-1)
(L_{i,2k-1}-L_{l,2k-1})
}}\right)^{1/2} & \cr
&&\cr
&\times \left(
{\prod_{i\ne 0=-k+2}^{k-2}(L_{i,2k-3}-L_{j,2k-2}-1)
\prod_{i\ne 0,l;i=-k+1}^{k-1}(L_{i,2k-1}-L_{j,2k-2}-1)
\over
{\prod_{i\ne 0,j; i=-k+1}^{k-2}(L_{i,2k-2}-L_{j,2k-2}-1)
(L_{i,2k-2}-L_{j,2k-2})
}}\right)^{1/2}  |M)_{-(l,2k-1)}^{-(j,2k-2)},& \cr
& \hskip 13cm k\in [2,\infty]. & (124) \cr
&&\cr
}
$$

The above transformation relations (116)-(124) were derived first for
$gl(n|1|n)$ from (68)-(74) and the supercommutation
relations. Therefore they give a representation of $gl(n|1|n)$ for any
$n$. An essential requirement, when passing to $n \rightarrow \infty
$, is given with the condition (113). It is straightforward to check
that $\V$ is invariant under the action of the generators. The rest of
the proof, which we skip, is rather similar to that of
Proposition 6, although technically it is more involved.

\vskip 1cm
\n
{\bf IV. CONCLUDING REMARKS}

\bigskip
We have constructed two classes of highest weight irreps of the
infinite-dimensional Lie superalgebra $\gl$. It should be noted that
the GZ representations are inequivalent to the
$C$-representations. More than that: the C-representations, being
highest weight irreps of $\2gl$, are not highest weight
representations of $\1gl$ and vice versa. Indeed, assume that the
$\1gl$ module $V([m])$ is also a highest weight $\2gl$ module with a
highest weight vector $y$. Then $y$ has to be a highest weight vector
of any of the subalgebras $gl(k|1|k-1+\t)$. Hence
Eqs. (54) and (55) have to hold for any $\t=0,1,\; k\in [1-\t;\infty]$.
Therefore $y\notin V([m])$ (see (87)).

Our primary interest in the present investigation is related to its
eventual applications in a generalization of the statistics in quantum
field theory. From this point of view our results are however very
preliminary. The first observation in this respect is that the algebra
(for definiteness) $\2gl$ is not large enough. It does not contain
important physical observables (like the energy-momentum of the
field $P^m$, see (8)), which are infinite linear combinations of the
generators of $\2gl$. In order to incorporate them one has to go to
the completed central extension $a(\infty|1|\infty)$
of $\2gl$ in a way similar as for the Lie algebra $gl_\infty$$^{41}$
or the Lie superalgebra $gl_{\infty|\infty}.^{20}$
This is only the first step. The next one
will be to determine those $\2gl$ modules $\V$, which can be extended
to $a(\infty|1|\infty)$ modules.

The most important and perhaps the most difficult step will be to
express the transformations of the $\2gl$ modules in terms of natural
for the QFT variables, namely via the creation and the annihilation
operators $a_i^\pm$ of $\2gl$, which are just its odd generators.$^4$
This is however not simple and, may be, even not necessary in the
general context of the representation theory. The physical state
spaces, the Fock spaces, have to satisfy several additional physical
requirements.$^{42}$ In particular any such space has to be generated
from the vacuum (the highest 
weight vector) by polynomials of the
creation operators, which are only a part of the negative root
vectors. This imposes considerable restriction on the physically
admissible modules. Hence in the applications one has to select first
the Fock spaces from all $\2gl$ modules and then study their
transformation properties under the action of the physically relevant
operators, in particular of the CAOs.

An additional problem is related to the circumstance that in QFT the
indices of the CAOs are not elements form a countable set. Therefore
as a test model one can try to consider first the $\2gl$ statistics in
the frame of a lattice quantum filed theory or locking the field in a
finite volume.

\vskip 1cm

\bigskip
\noindent
{\bf ACKNOWLEDGMENTS}

\smallskip
N.I.S. is grateful to Prof. M.D. Gould for the invitation to work in
his group at the Department of Mathematics in University of
Queensland. T.D.P. is thankful to Prof. S. Okubo for the kind
invitation to conduct a
 research under the Fulbright Program in the
Department of Physics and Astronomy, University of Rochester.
We wish to thank  Prof. Randjbar-Daemi for the kind 
hospitality at the High Energy Section of ICTP.

This work was supported by the Australian Research Council, by the
Fulbright Program of U.S.A., Grant No 21857, and by the Contract
$\Phi-416$ of the Bulgarian Foundation for Scientific Research.

\vskip 12pt
{\settabs \+  $^{11}$& I. Patera, T. D. Palev, Theoretical
   interpretation of the experiments on the elastic \cr
   %sample line,  see p. 232 of the Texbook.

\+ $^1$  & T.D. Palev, J. Math. Phys. {\bf 23}, 1778 (1982).\cr

\+ $^2$  & T.D. Palev, Czech. Journ. Phys. {\bf B32}, 680 (1982).\cr

\+ $^3$  & T.D. Palev, Czech. Journ. Phys. {\bf B29}, 91 (1979).\cr

\+ $^4$  & T.D. Palev, {\it A-superquantization}, Communication
           JINR E2-11942 (1978). \cr

\+ $^5$  & A.Ch. Ganchev and T.D. Palev, J. Math. Phys.
           {\bf 21}, 797 (1980).\cr

\+ $^6$  & H.S. Green, Phys. Rev. {\bf 90}, 270 (1953).  \cr

\+ $^7$ & N.N. Bogoljubov and D.V. Shirkov, {\it Introduction
          to the Theory of Quantized Fields},\cr
\+      & Moscow 1957 (English ed. Interscience Publishers, Inc.,
          New York, 1959)\cr

\+ $^{8}$ & E. Celeghini, T.D. Palev,  and M. Tarlini,
            Mod. Phys. Lett. {\bf B5}, 187 (1991).\cr

\+ $^{9}$ & T.D. Palev, J. Phys. A: Math. Gen. {\bf 26}, L1111 (1993)
            and hep-th/9306016. \cr

\+ $^{10}$ & L.K. Hadjiivanov, J. Math. Phys. {\bf 34}, 5476 (1993). \cr

\+ $^{11}$ & T.D. Palev and J. Van der Jeugt,
             J. Phys. A~: Math. Gen. {\bf 28}, 2605 (1995)
             and $q$-alg/9501020 .\cr

\+ $^{12}$ & T.D. Palev, Commun. Math. Phys. 
             {\bf 196}, 429 (1998) and $q$-alg/9709003.   \cr

\+ $^{13}$ & V.G. Kac,
            Lecture Notes in Math. {\bf 676}, 597 (Springer, 1979).\cr

\+ $^{14}$ &  S. Okubo, J. Math. Phys. {\bf 35}, 2785 (1994). \cr

\+ $^{15}$ & J. Van der Jeugt in  {\it New Trends in Quantum Field
             Theory (Heron Press, Sofia, 1996)}.\cr

\+ $^{16}$ & S. Meljanac, M. Milekovic and M. Stojic, {\it On
             parastatistics defines as} \cr
\+		   & {\it triple operator algebras} (q-alg/9712017).\cr

\+ $^{17}$ & T.D. Palev and N.I. Stoilova,
            J. Math. Phys. {\bf 38}, 2506 (1997) and
            hep-th/9606011. \cr

\+ $^{18}$ & F.D.M. Haldane, Phys. Rev. Lett. {\bf 67}, 937 (1991).\cr

\+ $^{19}$ & T.D. Palev and N.I. Stoilova, J. Phys. A~: Math. Gen.
            {\bf 27}, 977, 7387 (1994) \cr
\+		   & and hep-th/9307102, hep-th/9405125. \cr

\+ $^{20}$ & V.G. Kac and J.W. van der Leur, Ann. Inst. Fourier,
               Grenoble {\bf 37}, $\#$4, 99 (1987).\cr

\+ $^{21}$ & V.G. Kac and J.W. van der Leur, Advanced Series in
               Math. Phys. {\bf 7}, 369 (1988).\cr

\+ $^{22}$ & T.D. Palev, J. Math. Phys. {\bf 22}, 2127 (1981). \cr

\+ $^{23}$ & J. Van der Jeugt, J.W.B. Hughes, R.C. King,
              J. Thierry-Mieg, Commun. Algebra {\bf 18}, 3453 (1990).\cr

\+ $^{24}$ & H. Schlosser, Seminar Sophus Lie {\bf 3}, 15 (1993)\cr

\+ $^{25}$ & H. Schlosser, Beitr\"age zur Algebra and Geometry
               {\bf 31}, 193 (1994) \cr

\+ $^{26}$ & T.D. Palev, Funkt. Anal. Prilozh. {\bf 21,} N 3,
           85 (1987); Funct. Anal. Appl. {\bf 21,} 245 (1987) \cr
\+         & (English translation). \cr

\+ $^{27}$ & T.D. Palev, J. Math. Phys. {\bf 30}, 1433 (1989). \cr

\+ $^{28}$ & T.D.Palev and V.N.Tolstoy, Comm. Math. Phys.
             {\bf 141}, 549 (1991).\cr

\+ $^{29}$ & J. Van der Jeugt, J.W.B. Hughes, R.C. King,
             J. Thierry-Mieg, J. Math. Phys. {\bf 31}, 2278 (1990).\cr

\+ $^{30}$ & J.W.B. Hughes, R.C. King, J. Van der Jeugt,
             J. Math. Phys. {\bf 33}, 470 (1992).\cr

\+ $^{31}$ & T.D. Palev, Funkt. Anal. Prilozh. {\bf 23,} N 2,
           69 (1989); Funct. Anal. Appl. {\bf 23,} 141 (1989) \cr
\+         & (English translation). \cr

\+ $^{32}$ & J. Van der Jeugt, J. Math. Phys. {\bf 36}, 605 (1995).\cr

\+ $^{33}$  & T.D. Palev, N.I. Stoilova and J. Van der Jeugt,
              Comm. Math. Phys.
             {\bf 166}, 367 (1994).\cr

\+ $^{34}$ & T.D. Palev, Funkt. Anal. Prilozh. {\bf 24,} N 1,
           69 (1990); Funct. Anal. Appl. {\bf 24,} 72 (1990) \cr
\+         & (English translation). \cr

\+ $^{35}$ & T.D. Palev, J. Math. Phys. {\bf 31}, 579 (1990) and
            {\bf 31}, 1078 (1990). \cr

\+ $^{36}$ & V.V. Serganova, Math. USSR Izv. {\bf 24}, 359 (1985). \cr

\+ $^{37}$ & D.A. Leites, M.V. Saveliev and V.V. Serganova,
             Serpukhof preprint 85-81 (1985).\cr

\+ $^{38}$ & J.W. Van der Leur, Cotragradient Lie superalgebras of
              finite growth, Utrecht thesis (1985).\cr

\+ $^{39}$ & I. Penkov and V. Serganova,
           Indag. Math. {\bf 3}, 419 (1992). \cr

\+ $^{40}$ & V.G. Kac and M. Wakimoto,
           Progress in Math.  {\bf 123}, 415 (1994). \cr

\+ $^{41}$ & V.G. Kac and V.G. Peterson,
           Proc. Natl. Acad. Sci. USA {\bf 78}, 3308 (1981). \cr

\+ $^{42}$ & T.D. Palev, J. Math. Phys. {\bf 21}, 1293 (1980). \cr

\end